\PassOptionsToPackage{table}{xcolor}
\documentclass[astrosymb, twocolumn,  ]{aastex631} %,linenumbers
\usepackage[table]{xcolor}
\usepackage{subfigure}
\usepackage{mathtools}
\usepackage{hyperref}
\received{18 June 2025}
\revised{1 October 2025}
\accepted{16 October 2025}
\submitjournal{\apj}
\shorttitle{}
\shortauthors{}
\graphicspath{{./}{./}}
\begin{document}
\title{Active Dwarf Galaxy Database II: Connections between Host Galaxy Properties and Black Hole Accretion Signatures}

[\author[0000-0003-3986-9427]{Erik J. Wasleske}
\affiliation{Department of Astronomy, University of Illinois at Urbana-Champaign, 1002 West Green Street,  Urbana, IL 61801, USA}
\affiliation{Department of Physics and Astronomy, Washington State University, Pullman, WA 99163, USA}
\email{wasleske@illinois.edu}]

\author[0000-0003-4703-7276]{Vivienne F. Baldassare}
\affiliation{Department of Physics and Astronomy, Washington State University, Pullman, WA 99163, USA}

\author[0000-0003-3574-2963]{Christopher M. Carroll}
\affiliation{Department of Physics and Astronomy, Washington State University, Pullman, WA 99163, USA}

%%%%%%%%%%%%%%%%%%%%%%%%%%%%%%%%%%%%%%%%%%%%%%%%%%%%%%%%%%%
%%
%%      Abstract
%%
%%%%%%%%%%%%%%%%%%%%%%%%%%%%%%%%%%%%%%%%%%%%%%%%%%%%%%%%%%%
\begin{abstract}

We investigate the connection between accretion signatures and host galaxy properties in the context of how active dwarf galaxies are identified. 
We use the database constructed in \cite{wasleske2024} which contains dwarf galaxies that were selected as active galaxies by optical spectroscopy, infrared colors, X-ray brightness, and photometric variability. Multi-wavelength archival data was used to consistently apply all of these methods to every galaxy within this compiled set. The cross application of these methods resulted in a diversity of sub-populations identified as active by some set of these techniques. In this paper, we estimate host galaxy properties from spectral energy distribution models. We connect the active galactic nuclei (AGN) signatures to our estimated host galaxies' properties using statistical dimensionality reduction methods. We find that dwarf AGN selected by infrared colors are the most distinct population, with the highest star formation rates and lowest stellar masses. We also find some other key population differences, such as the broad line AGN having significantly higher AGN luminosities. X-ray and variability selected AGN have higher average star formation rates than those selected with optical narrow line spectroscopic diagrams.   
Our connections to the host galaxy parameters potentially point to the sub-populations representing different epochs of the evolution of accretion. 

\end{abstract}


%%%%%%%%%%%%%%%%%%%%%%%%%%%%%%%%%%%%%%%%%%%%%%%%%%%%%%%%%%%
%%
%%      Introduction
%%
%%%%%%%%%%%%%%%%%%%%%%%%%%%%%%%%%%%%%%%%%%%%%%%%%%%%%%%%%%%
\section{Introduction} \label{sec:intro}

The correlation between the mass of host galaxies and their central massive black hole (MBH) has been well studied \citep{ferrarese2000, gebhardt2000,kormendy2013}. These MBHs can affect host galaxy properties such as morphology and star-formation (SF; \citealt{cattaneo2009}). Conversely, the gas, dust, and SF within the galaxy can mask the emission from the active MBH. This adds to the difficulty of compiling a complete census of active galaxies.

There has been extensive work done to establish a set of methods to identify active MBHs in the center of galaxies (e.g., \citealt{ho1997, ulrich1997, berk2004, just2007, desroches2009, assef2010, stern2012}). Each method selects MBHs based on different emission signatures stemming from the accretion of material within the active galactic nucleus (AGN).
The canonical picture of an AGN comprises the central accreting black hole (BH) surrounded by an accretion disk, with an X-ray corona, enveloped by a dusty torus with polar jets \citep{antonucci1993}. With each type of observation, different selection biases are introduced from limitations in observing the radiation from these components.

Spectroscopic methods (e.g., \citealt{baldwin1981, kewley2006, shirazi2012}) use narrow emission line ratios to separate signatures of accreting BHs from stellar processes. 
These diagrams can also distinguish Seyfert galaxies from low-ionization emission line ratio (LINER) type galaxies. For dwarf galaxies ($M_* \leq 10^{9.5}M_\odot$) which contain less massive central BHs compared to more massive galaxies, their lower accretion power may not outshine the stellar processes. Additionally, the lower host galaxy metallicities alter the value of the line ratios to move active galaxies into the star forming region \citep{groves2006, reines2013}. This increases the potential for active BHs in dwarf galaxies to be misclassified. The models of \cite{cann2019} showed that with decreased mass, a galaxy shifts out of the AGN region of the standard BPT diagram \citep{baldwin1981} as the lower mass BHs have more extended partially ionized zones. Broad emission lines---a signature of Doppler broadening due to the close proximity of gas to the central BH--are less frequently observed in dwarf candidates \citep{reines2013} due to lower BH masses.

Bright X-ray emission has long been an observed feature of AGN \citep{brandt2005, greene2007, desroches2009, lemons2015, plotkin2016, birchall2020}. The strength of this X-ray emission is known to scale with BH mass \citep{merloni2003}. Although X-rays are less susceptible to the effects of obscuration, one must rule out other possible sources of X-ray emission. {X-ray binaries and supernovae remnants serve as a sources of X-ray emission, leading to a tight correlation of hard X-rays to star formation rate in star forming galaxies \citep{ranalli2003}. In turn, these phenomena must be ruled out to determine if the X-ray emission is from accretion.}
This {source of contamination} is more prevalent when {searching for} weaker X-ray emissions of {active BHs in} dwarf galaxies {which can be masked by enhanced emission from low metallicity stellar populations \citep{lehmer2021}}.

Infrared (IR) colors measured from the Wide-field Infrared Survey Explorer (WISE; \citealt{wright2010}) have been used to identify AGN \citep{jarrett2011,stern2012,assef2018,hviding2022} as the WISE wavelengths are sensitive to the obscuring torus of the AGN. {These WISE IR colors are biased towards AGN that dominate the host galaxy stellar emission.}
However, these selection techniques are heavily influenced by dust and star formation, {as extreme starburst activity within galaxies can mimic AGN within these selections \citep{satyapal2018}}. \cite{hainline2016} found the majority of {optically selected} dwarf galaxies{,} with active signatures from their narrow emission lines{,} are not selected using these IR color criteria.

{The intrinsic variability of AGN has been used as a tool to identify these objects within time domain data with good success (e.g \citealt{schmidt2010, macleod2011, baldassare2018, burke2022}) and is of particular interest as new time domain surveys come online. Recent works have searched for correlations of variability to characteristics of the AGN (eg. characteristic dampening time to black hole mass \citealt{burke2021b, wang2023}) 
However, technical survey details can limit the detection of variability. The baseline of observations, cadence, and gaps in data can impact the modeling of the varying stochastic signal \citep{kozlowski2017}. Additionally, as variability is investigated often from optical imaging, handling host galaxy contribution requires great care to accurately characterize variations from the AGN's emission, especially if it is of lower power.  }

The inherent issues of each technique listed above lead to an incomplete understanding of the population of active dwarf galaxies. To collect the most complete and correct sample of active dwarf galaxies, one must implement multi-wavelength observations. Accurate populations can be used to constrain seeding mechanisms of central galactic BHs and the evolutionary relationships with their host galaxy (\citealt{greene2020} and references therein).

In \cite{wasleske2024} we compiled a set of 733 active dwarf galaxies  that were initially identified using optical spectroscopy, X-ray, infrared (IR), and optical photometric variability. We collect available archival data and applied a uniform set of selection techniques to this database. After the application of methods using optical line diagnostic diagrams, X-ray photometry, IR colors, and photometric variability, identifying a sub-sample of galaxies selected as AGN from each method. We then cross applied these methods to these subsets to investigate the overlapping populations identified by these methods. This resulted in a lacking agreement between these subsets. 

In this paper, we investigate the biases in these selections and search for a parameter space that distinguishes these populations from another.
We first estimate host galaxy properties and connect them to selection methods. We model the spectral energy distributions (SEDs) to investigate the distribution of host parameters within the AGN subsamples.
We then use t-distributed stochastic neighbor embedding (t-SNE, \citealt{maaten2008}) as a clustering tool to group galaxies based on a combination of the host galaxy parameters, which in turn allows us to decouple galaxies identified with multiple AGN signatures. %This is turn will help us decouple those galaxies that were identified from multiple AGN signatures

This paper is organized as follows: Section \ref{sec:data} we compile additional photometry for our analysis;
Section \ref{sec: Methods} describes our construction of spectral energy distribution (SED) models using this photometry and our application of t-SNE clustering;
Section \ref{sec:results} describes the results drawn from the selection subsamples and the t-SNE clusters; Section \ref{sec:discussion} considers the implications of these results and the influence host galaxy properties have on the selection of active dwarf galaxies. We assume a $\Lambda$CDM cosmology with parameters $h=0.7$, $\Omega_m=0.3$ and $\Omega_\lambda=0.7$ \citep{spergel2007}.

%%%%%%%%%%%%%%%%%%%%%%%%%%%%%%%%%%%%%%%%%%%%%%%%%%%%%%%%%%%
%%
%%      Data
%%
%%%%%%%%%%%%%%%%%%%%%%%%%%%%%%%%%%%%%%%%%%%%%%%%%%%%%%%%%%%
\section{Data} \label{sec:data}

We start with the resulting active dwarf galaxy database (ADGD) from (\citealt{wasleske2024}; hereafter W24), using these positions to collect more accurate photometry and results where needed. %This contains compiled photometry from CXO/XMM X-ray sources and WISE, as well as measured optical emission line fluxes.

Our sample is comprised of the resulting 702 active dwarf galaxies identified from a uniform application of AGN-selection techniques.
To accurately model and constrain the SEDs of our sample, we compiled additional photometry to supplement W24.
This involves collecting optical photometry as no optical photometry was used in W24, and additional IR and UV data.

%%%
% SDSS:
To collect SDSS \textit{ugriz} observations from DR 17 \citep{abdurrouf2022}, we use \texttt{astroquery} to search for matches within a 3" radius. We use the \texttt{cModelMag} values for these five bands of the closest source, restricting to only clean photometry (ie. only keep photometry values for whose \texttt{`CLEAN'} value is 1). Extinctions of these magnitudes were used to correct \texttt{cModelMag} to calculate the dereddened magnitude in each band. No values were found for 46 of our objects.

We use the position of the SDSS source to query the rest of the photometry discussed here. For the 46 objects with no SDSS photometry, we used their positions from the ADGD to query the rest of the photometry listed below. {Of the 46 objects with no SDSS photometry, 20 have photometric values in at least one band of Pan-STARRS DR2 \citep{chambers2016,flewelling2020}, ten of which have values in each \textit{grizy} filter. For consistency, we opt to not include this set of Pan-STARRS photometry into our analysis.}

% WISE:
Furthermore, we collect photometry from the Wide-Field Infrared Survey Explorer (WISE, \citealt{wright2010}) AllWISE Source Catalog (\citealt{wright2019}) for each galaxy{, within a 3$''$ radius,} across the four infrared bands to supersede the use of the WISE All-Sky Release catalog in W24.
We compile the profile-fitting photometry magnitudes for these four bands (W1, W2, W3, W4 bands with wavelengths 3.4, 4.6, 12 and 22 $\mu\text{m}$ respectively). W1, W2, and W4 data was available for 694 of these objects, with W3 data available for all but one of these 694 objects. 
% UKDSS
For UKIDSS/UKIRT \citep{lawrence2007}, we use \texttt{astroquery} to first search the Large Area Survey (LAS) in a Y, H, J and K bands. We collected Petrosian magnitudes for each of these bands from the \texttt{UKIDSSDR11PLUS} catalog search within 3" of the galaxy positions. This data was available for 302 objects within the ADGD. Petrosian magnitudes should have good agreement with the \texttt{cModelMag} of SDSS \citep{strauss2002}. 

% GALEX
Additionally, we searched for available Galaxy Evolution Explorer (GALEX) near($1750-2800$ \AA) and  far(1350-1750 \AA) ultra-violet fluxes within 3" of each position. We used \texttt{astroquery} to search the GUVcatAIS GR 6+7 catalog (SourceRevised catalog of GALEX UV sources (GUVcat\_AIS GR6+7), \citealt{bianchi2017}). NUV values was available for 504 of the objects and FUV values for 442 of the objects. 

{The X-ray photometry is taken directly from the ADGD as found by the Chandra X-ray Observatory and XMM-Newton analysis in W24.}
We compiled all of these values as inputs for our spectral energy distribution modeling described below.

%%%%%%%%%%%%%%%%%%%%%%%%%%%%%%%%%%%%%%%%%%%%%%%%%%%%%%%%%%%
%%
%%      Methods
%%
%%%%%%%%%%%%%%%%%%%%%%%%%%%%%%%%%%%%%%%%%%%%%%%%%%%%%%%%%%%
\section{Methods} 
\label{sec: Methods}

We created SED models to estimate the host galaxy properties. \cite{ramos_padilla2021} used SED models in unison with machine learning algorithms to test to standard AGN model and its assumption of classification based on viewing angle, showing the utility of this combination. In the following, we first discuss the construction of our SED models by illustrating the modules and parameter grids used. After, we list the machine learning algorithms implemented to draw the AGN classification-host galaxy parameter connection.

\subsection{Spectral Energy Distribution Modeling}
\label{subsec: methods - SED}

\begin{figure}[h]
  \centering
    \subfigure{\includegraphics[width=0.475\textwidth]{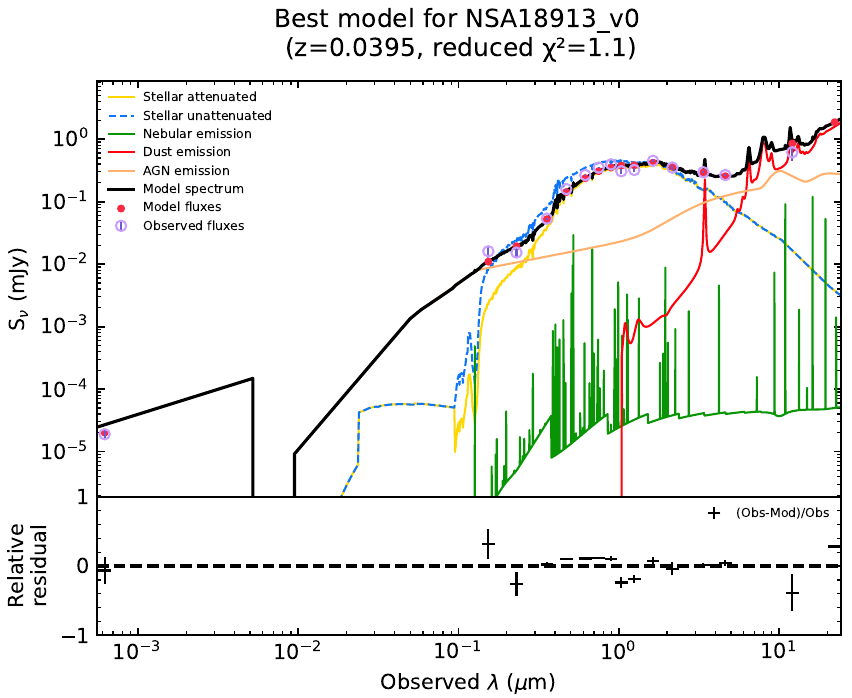}}%\quad
    \hspace{1em}% Space between image A and B
    \subfigure{\includegraphics[width=0.475\textwidth]{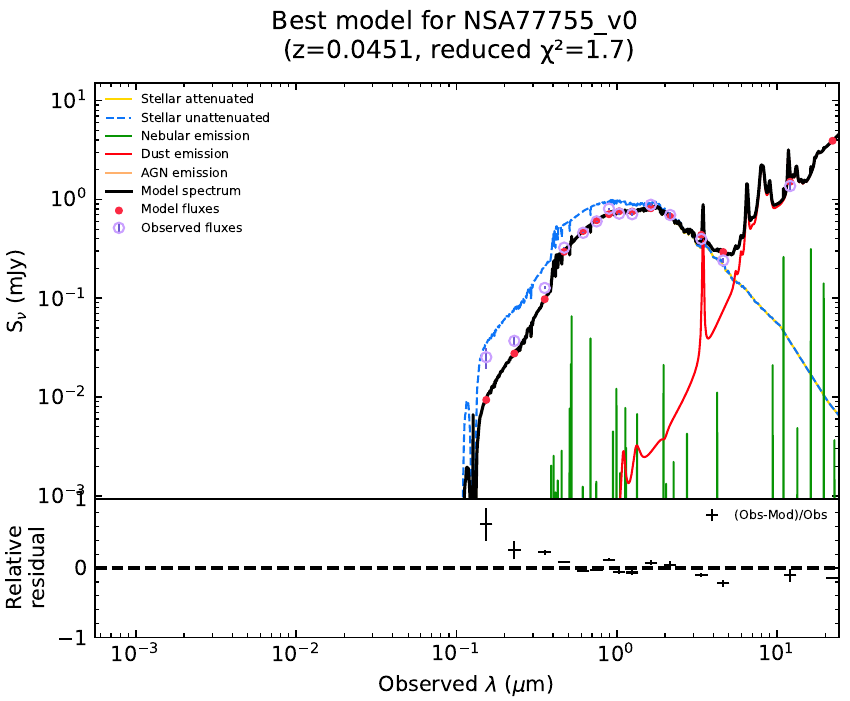}}
  \caption{
  SED models for NSA 18993(top) and NSA 77755(bottom). NSA 18993 was selected as an AGN via the BPT and [\ion{He}{2}] diagrams, as a Seyfert on the  [\ion{S}{2}] diagram, and from its broad H$\alpha$ and X-ray emissions. In contrast, NSA 77755 was only selected based on optical variability.
  }
  \label{fig: SED examples}
\end{figure}

We used \texttt{X-CIGALE V2022.0} \citep{yang2022} to model the SEDs of the galaxies in our sample. \texttt{CIGALE} \citep{noll2009, boquien2019} is a Python-based code created to efficiently model the FUV-to-radio spectrum of galaxies. It estimates galaxy properties by using multi-wavelength photometric observations of galaxies to constrain models of stellar, dust attenuation, nebular and AGN emission. \texttt{CIGALE} uses energy balance principles to give results for large samples of galaxies within much shorter computational time than solving the 3-dimensional radiative transfer equation whilst still having high accuracy on parameter estimates. The properties estimated from these SEDs use Bayesian approach by weighing all models by their goodness-of-fit. These methods includes observational uncertainties and the degeneracy between the parameters. \texttt{X-CIGALE} builds upon the \texttt{CIGALE}  by including X-ray photometry fitting and a model of polar dust in AGNs, making this fitting code ideal for AGN.

We present the modules and parameters used in our implementation of \texttt{X-CIGALE} to generate SED models of the galaxies in the ADGD. We base our parameter grid off of the one used in \cite{wasleske2023}, with modifications listed below and outlined in Table \ref{tab:CIGALE parms}. 
\begin{table*}[t]
    \centering
    \textbf{\texttt{X-CIGALE} Parameters\\}
    \begin{tabular}{|c  c c  |}
\hline
Parameter & Value & Description
\\
\hline

\hline
\multicolumn{3}{|c|}{\textbf{Star Formation History: sfhdelayed} }
\\
$\tau_{\text{main}}$ & 100, 500, 1000, 5000 Myrs& e-folding time for main stellar population model.
\\
$\text{age}_{\text{main}}$ & 1000, 2000, 5000 Myrs & Age of the main stellar population in the galaxy.
\\

\hline
\multicolumn{3}{|c|}{\textbf{Single-age Stellar Population: bc03} \citep{bruzual2003}}
\\
IMF & 1 & \cite{chabrier2003} initial mass function
\\
metallicities & 0.0001, 0.0004, 0.004, 0.008, 0.02 & Range for stellar population
\\
Separation Age & 10 Myrs & Differentiate old \& young stellar populations 
\\

\hline
\multicolumn{3}{|c|}{\textbf{Nebular Model: nebular} }
\\
$\text{log}\:\text{U}$ & -2.0 & Ionization parameter.
\\
$\text{z}_{\text{gas}}$ & 0.004, 0.008, 0.014, 0.033 & Range for gas metallicity. 
\\

\hline
\multicolumn{3}{|c|}{\textbf{Dust Attenuation: dustatt\_modified\_starburst} (Modified \citealt{calzetti2000} Attenuation Law)}
\\
Extinction law & MW, LMC & For attenuating the emission line flux.
\\

\hline
\multicolumn{3}{|c|}{\textbf{Dust Emission: dl2014} \citep{draine2014}}
\\
$\text{Q}_{\text{PAH}}$ & 2.5 & Mass fraction of PAH.
\\
$\text{U}_{\text{min}}$ & 1.0 & Minimum radiation field.
\\
$\alpha$ & 2.0 & Powerlaw slope dU/dM propto $U^{\alpha}$
\\
$\gamma$ & 0.1 & Fraction illuminated from $U_{\text{min}}$ to $U_{\text{max}}$.
\\

\hline
\multicolumn{3}{|c|}{\textbf{AGN model: skirtor2016} \citep{stalevski2016}}
\\
$\text{frac}_{\text{AGN}}$ & 0.0, 0.1, 0.25, 0.5, 0.6, 0.7, 0.8, 0.9 & Fraction of AGN torus to the IR total luminosity.
\\
E(B-V) & 0.0, 0.025, 0.05, 0.075, 0.10, 0.125, 0.15, 0.175 & Excess Color in polar direction.
\\
i & 10, 30, 50, 70$^{\circ}$  & viewing angle with respect to the AGN.
\\

\hline
\multicolumn{3}{|c|}{\textbf{X-ray} (from skirtor2016 AGN)}
\\
$\alpha_{\text{OX}}$ & -2.0, -1.75, -1.5, -1.25, -1.0 & Power law slope defined in \cite{just2007}.
\\
$\Gamma$ & 1.8 & Photon index of the intrinsic AGN x-ray spectrum
\\

\hline
\hline

\end{tabular}
    \caption{Grid parameters for the \texttt{X-CIGALE} SED modeling.}
    \label{tab:CIGALE parms}
\end{table*}

First we include a delayed star formation history module (\texttt{sfhdelayed}) to adequately account for early- and late-type galaxies \citep{boquien2019} with e-folding times of 100, 500, 1000, or 5000 Myr.  
Moreover, the stellar model of \cite{bruzual2003} is suited to fit stellar populations with ranging metallicities and ages. We allow for fits with metallicities of 0.0001 -- 0.02 for a Chabrier initial mass function \citep{chabrier2003}.  
%Next, the nebular emissions is modeled by BLANK.
We include the modified dust-attenuated starburst module from \cite{calzetti2000}. 

For our AGN contribution, we used the \texttt{SKIRTOR} \citep{stalevski2012,stalevski2012b,stalevski2016} module. This module implements a two-phase dusty torus of low density medium filled with high density clumps. We set the inclination angle to the set of 10\textdegree, 30\textdegree, 50\textdegree, or 70\textdegree\: and the AGN fraction to values from 0.0 to 0.9.

The dust contribution is modeled by \texttt{dl2014}. This module is based of the dust model of M31 \citep{draine2007}, which assume dust comprised of carbonaceous grains and amorphous silicate grains and variable polycyclic aromatic hydrocarbon abundances. The size distribution of the grains is consistent with the extinction and IR emission of diffuse interstellar medium within the solar neighborhood.

We only kept those objects whose SED model has $0.5 < \chi_{r}^2 \leq 10$. This left 461 galaxies to analyze. The number of these galaxies within each sub-sample is given in the first column of Table \ref{tab:subsample parameters}. {The preliminary values for host galaxy properties that can be derived from our data (i.e. the SFR derived in W24 using corrected H$\alpha$ measured from spectrum) were not used to inform the priors for our SED modeling. As both values taken from derived relations the \texttt{X-CIGALE} modeling have systematic uncertainty, we chose to keep our SED models agnostic to previous results.}

From these models we collect Bayesian estimates of host galaxy dust and stellar luminosity, the stellar metallicities, and the AGN luminosity. Example models are shown in Figure \ref{fig: SED examples}. We will be using redshift and {host galaxy stellar} mass from W24 alongside the estimated values of stellar metallicity, $L_\text{dust}$, $L_\text{AGN}$, $L_\text{stellar}$ and E(B-V) for the input to all machine learning algorithms explored below. $L_\text{AGN}$ is sum of the AGN disk and dust re-emitted luminosity taken from the \texttt{SKIRTOR} module. $L_\text{dust}$ is the estimated dust luminosity of the (\texttt{dl2014}, \citealt{draine2014}) module from an energy balance, integrated over the full SED.  $L_\text{stellar}$ is estimated from the \texttt{bc03} module. 
We give a summary of the mean values of these parameters for each sub-sample of the ADGD in Table \ref{tab:subsample parameters}.

%\begin{landscape}
%\begin{rotatetable}
%\movetabledown=20mm
%\movetabledown=
%\begin{sidewaystable*}[b]
%\newpage

%\begin{rotatetable*}%[h]
%\begin{deluxetable*}
%\begin{table*}
\movetabledown=5cm
\begin{table*}
\begin{rotatetable*}
    \centering
    \textbf{Median Sub-Sample Host Parameter values\\}
\begin{tabular}{c | c c c c c c c c c }
\hline
Selection Method & Redshift & Mass & SFR & 
$\text{log}(L_{\text{AGN}})$ & $\text{log}(L_{\text{dust}})$ & $\text{log}(L_{\text{stellar}})$ & \textbf{Stellar }Metallicities & $E(B-V)$ & $\text{f}_{AGN}$\\

 &   &  [$\text{M}_\odot$] & [$\text{M}_\odot$ / yr] &
 [erg/s] & [erg/s] & [erg/s] &   &   & \\
%\hline
%    &   \%   & \% & \%  \\
\hline
\hline

\begin{tabular}[c]{@{}l@{}}BPT (AGN+Comps)\\ Total: 197\end{tabular}& 
$0.032_{-0.011}^{+0.02}$&
$9.33_{-0.29}^{+0.12}$&
$0.003_{-0.003}^{+0.138}$ &
$40.98_{-2.15}^{+1.72}$&
$42.68_{-0.33}^{+0.51}$&
$43.18_{-0.35}^{+0.42}$&
$0.0008_{-0.0005}^{+0.0059}$&
$0.0877_{-0.0078}^{+0.0146}$&
$0.0019_{-0.0}^{+0.1782}$
\\
\hline

\begin{tabular}[c]{@{}l@{}}{[\ion{O}{1}]} Seyferts\\ Total: 181\end{tabular} &
$0.034_{-0.012}^{+0.039}$&
$9.21_{-052}^{+0.24}$&
$0.008_{-0.007}^{+0.352}$&
$40.19_{-1.83}^{+2.3}$&
$42.66_{-0.42}^{+0.62}$&
$43.13_{-0.36}^{+0.45}$&
$0.0004_{-0.0002}^{+0.0048}$&
$0.0875_{-0.0014}^{+0.0034}$&
$0.0002_{-0.0}^{+0.1034}$
\\
\hline

\begin{tabular}[c]{@{}l@{}}{[\ion{O}{1}]} LINERS\\ Total: 23\end{tabular} & 
$0.028_{-0.008}^{+0.016}$&
$9.31_{-0.32}^{+0.15}$&
$0.003_{-0.003}^{+0.177}$&
$39.28_{-1.97}^{+2.53}$&
$42.48_{-0.31}^{+0.54}$&
$42.92_{-0.32}^{+0.38}$&
$0.0004_{-0.0002}^{+0.0065}$&
$0.0875_{-0.0}^{+0.0015}$&
$0.0001_{-0.0}^{+0.0573}$
\\
\hline

\begin{tabular}[c]{@{}l@{}}{[\ion{S}{2}]} Seyferts\\ Total:190\end{tabular} & 
$0.036_{-0.014}^{+0.03}$&
$9.24_{-0.50}^{+0.21}$&
$0.005_{-0.004}^{+0.13}$&
$40.13_{-1.96}^{+2.25}$&
$42.67_{-0.34}^{+0.47}$&
$43.13_{-0.33}^{+0.39}$&
$0.0004_{-0.0002}^{+0.004}$&
$0.0875_{-0.0013}^{+0.0025}$&
$0.0002_{-0.0}^{+0.0917}$
\\
\hline

\begin{tabular}[c]{@{}l@{}}{[\ion{S}{2}]} LINERS\\ Total: 39\end{tabular} & 
$0.028_{-0.006}^{+0.009}$&
$9.24_{-0.42}^{+0.17}$&
$0.003_{-0.002}^{+0.02}$&
$39.70_{-2.89}^{+1.8}$&
$42.44_{-0.33}^{+0.23}$&
$42.84_{-0.39}^{+0.33}$&
$0.0004_{-0.0002}^{+0.006}$&
$0.0875_{-0.0023}^{+0.0004}$&
$0.0000_{-0.0}^{+0.0341}$
\\
\hline

\begin{tabular}[c]{@{}l@{}}{[\ion{He}{2}]} (AGN+Comps)\\ Total: 241\end{tabular} & 
$0.031_{-0.001}^{+0.021}$&
$9.29_{-0.44}^{+0.16}$&
$0.003_{-0.003}^{+0.08}$&
$40.65_{-2.24}^{+1.79}$&
$42.61_{-0.37}^{+0.45}$&
$43.1_{-0.35}^{+0.43}$&
$0.0005_{-0.0002}^{+0.0059}$&
$0.0875_{-0.0115}^{+0.0029}$&
$0.0004_{-0.0}^{+0.1036}$
\\
\hline

\begin{tabular}[c]{@{}l@{}}Broad-line\\ Total: 21\end{tabular} &
$0.047_{-0.018}^{+0.05}$&
$9.26_{-0.23}^{+0.18}$&
$0.109_{-0.108}^{+1.458}$&
$43.08_{-1.5}^{+0.62}$&
$43.18_{-0.61}^{+0.58}$&
$43.53_{-0.44}^{+0.49}$&
$0.0012_{-0.0008}^{+0.0071}$&
$0.0877_{-0.073}^{+0.0617}$&
$0.1919_{-0.0}^{+0.6894}$
\\
\hline

\begin{tabular}[c]{@{}l@{}}WISE IR\\ Total: 133\end{tabular} &
$0.055_{-0.029}^{+0.025}$&
$8.83_{-0.52}^{+0.44}$&
$0.144_{-0.138}^{+1.143}$&
$41.10_{-2.53}^{+1.97}$&
$43.07_{-0.69}^{+0.65}$&
$43.41_{-0.71}^{+0.54}$&
$0.0003_{-0.0002}^{+0.003}$&
$0.0876_{-0.0011}^{+0.0229}$&
$0.0002_{-0.0}^{+0.2335}$
\\
\hline

\begin{tabular}[c]{@{}l@{}}X-ray (AGN)\\ Total: 28\end{tabular} &
$0.031_{-0.016}^{+0.016}$&
$9.28_{-0.25}^{+0.14}$&
$0.007_{-0.007}^{+0.375}$&
$41.51_{-3.08}^{+1.47}$&
$42.72_{-0.27}^{+0.63}$&
$43.15_{-0.32}^{+0.52}$&
$0.0006_{-0.0004}^{+0.011}$&
$0.0875_{-0.0462}^{+0.029}$&
$0.000_{-0.0}^{+0.2054}$
\\
\hline

\begin{tabular}[c]{@{}l@{}}Variability\\ Total: 84\end{tabular} & 
$0.034_{-0.01}^{+0.02}$&
$9.24_{-0.68}^{+0.18}$&
$0.006_{-0.005}^{+0.112}$&
$40.43_{-2.37}^{+2.01}$&
$42.68_{-0.44}^{+0.52}$&
$43.1_{-0.32}^{+0.47}$&
$0.0004_{-0.0002}^{+0.0066}$&
$0.0875_{-0.0146}^{+0.0015}$&
$0.000_{-0.0}^{+0.0654}$
\\

\hline

\end{tabular}

%
%\caption{ The median values of the host galaxy parameters estimated from the SED modeling. The total number of objects is the remaining size of the sub-sample once the $\chi_{r}^2$ cut was made. Upper and lower bounds given the width of 1 $\sigma$ within each distribution.  }

%\end{center}
\caption{
 The median values of the host galaxy parameters estimated from the SED modeling. The total number of objects is the remaining size of the sub-sample once the $\chi_{r}^2$ cut was made. Upper and lower bounds given the width of 1 $\sigma$ within each distribution. 
}
\label{tab:subsample parameters}
\end{rotatetable*}
\end{table*}

%\end{deluxetable*}
%\end{rotatetable*}
%\end{table*}
%\end{sidewaystable*}
%\end{rotatetable}
%\end{landscape}

%%
\subsection{Dimensionality Reduction of Host Galaxy Parameter Space}
\label{subsec: methods - tSNE}
To search for connections between AGN selection method and host galaxy properties, we adopted the t-distributed stochastic neighbor embedding (t-SNE; \citealt{maaten2008}) dimensionality reduction algorithm to uncover potential patterns and groupings in the data.
This algorithm is based on the stochastic neighbor embedding technique \citep{hinton2002} which uses the sum of Kullback-Leibler divergences as a cost function while using a Student-t distribution instead of a Gaussian to compute the similarity of pair of point in the low-dimensional space. In addition to the construction of the cost function, the user has the ability to adjust the perplexity of this mapping, a hyperparameter that defines a smooth measure of the number of neighbors for each point within the low-dimensional space. This algorithm is fairly sensitive to changes in perplexity. 
This method retains the local structure and reveals the global structure of the data within its low-dimensional map. 

\begin{figure}[h]
  \centering
    \subfigure{\includegraphics[width=0.5\textwidth]{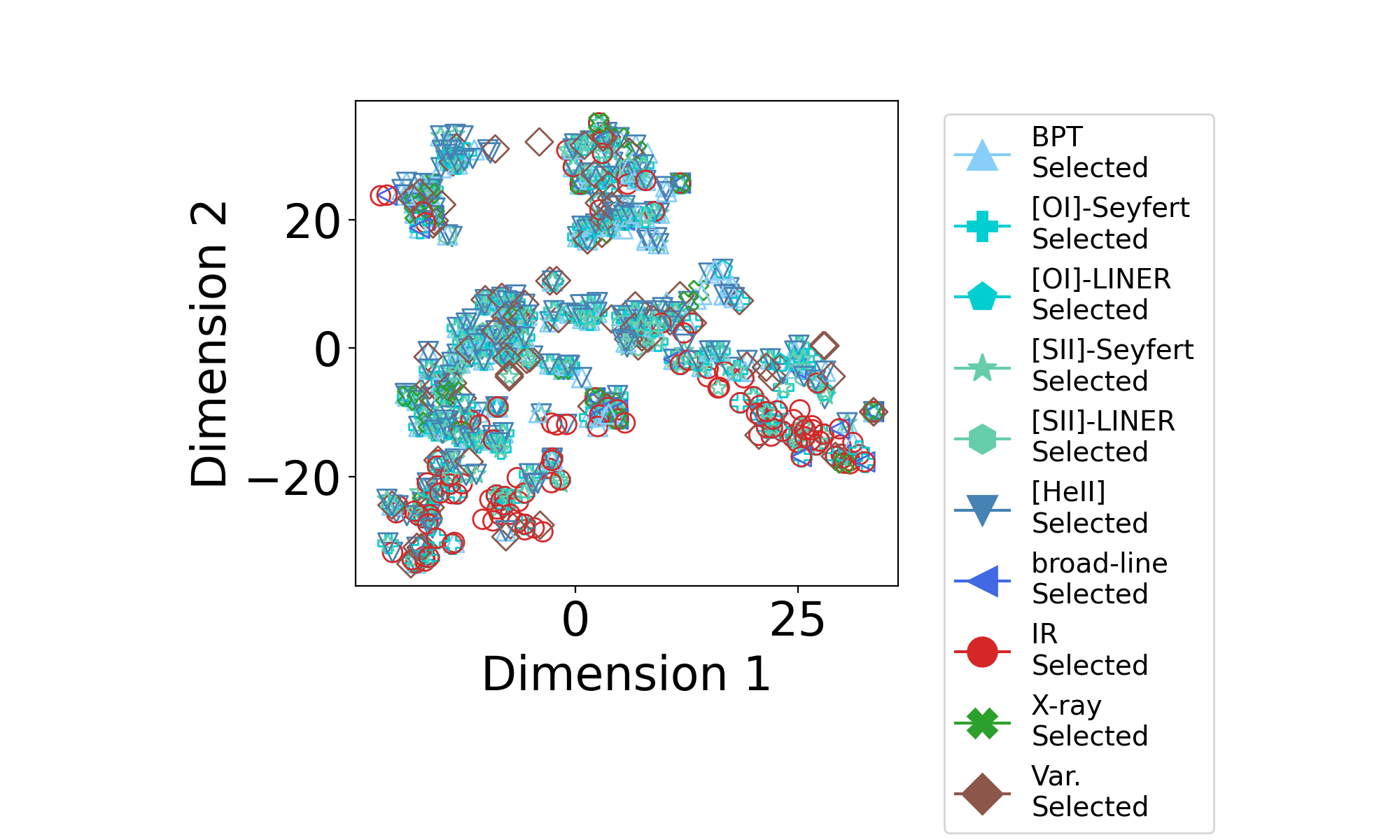}}%\quad
    \hspace{1em}% Space between image A and B
    \subfigure{\includegraphics[width=0.5\textwidth]{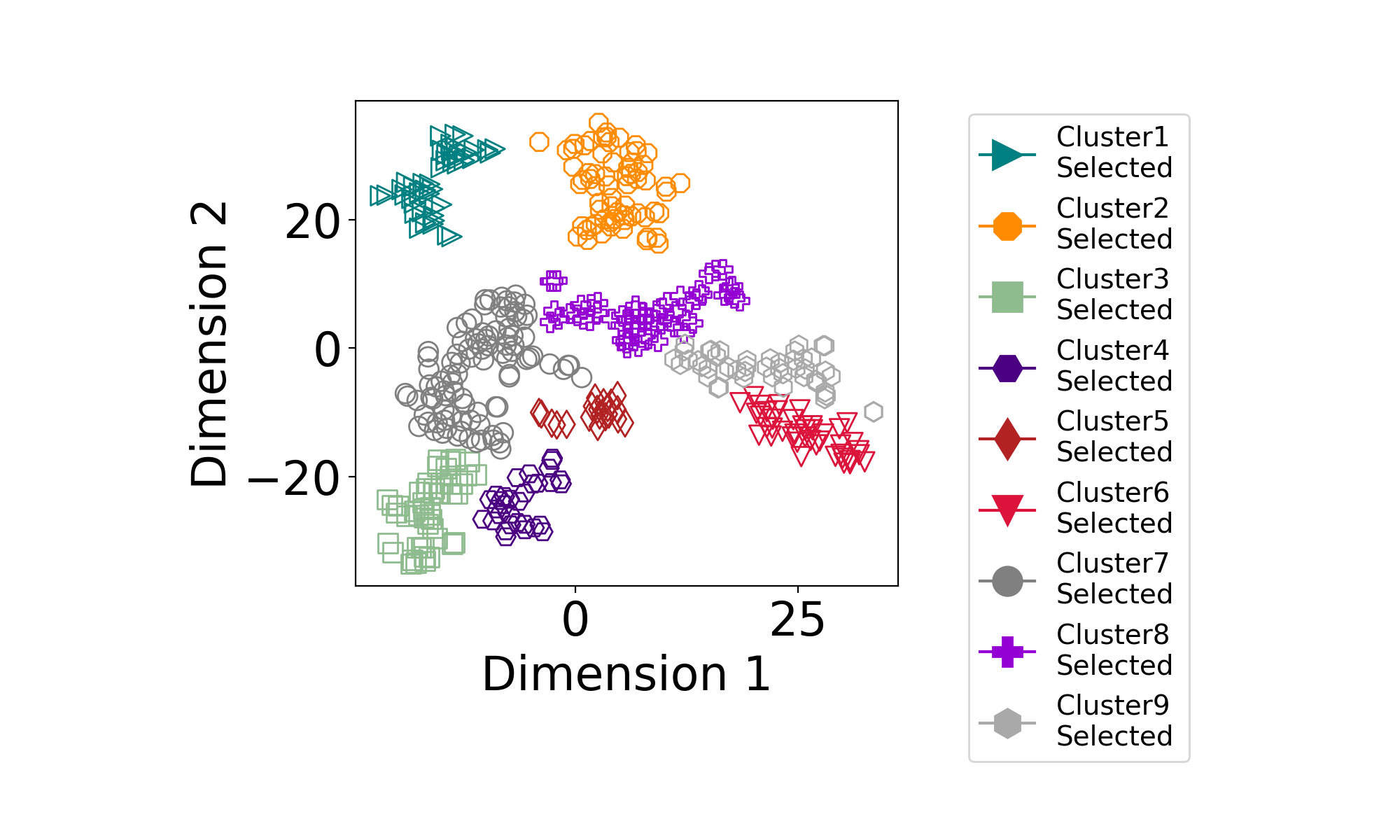}}
  \caption{t-SNE clustering with a perplexity of 20, colored by AGN selection method (\textit{top}) and after re-binning them into Clusters (\textit{bottom}). }
  \label{fig: tSNE clusters}
\end{figure}

We input the redshift, {host galaxy stellar} mass, and Bayesian estimates of SFR, stellar metallicity, dust luminosity $L_{\text{dust}}$, AGN luminosity $L_{\text{AGN}}$, stellar luminosity $L_{\text{stellar}}$ and excess E(B-V) color from the SED models. 
Figure \ref{fig: tSNE clusters} shows the results of this algorithm. We re-bin objects based on their position in the lower 2d dimensional map shown. We note that this method is fundamentally a data visualization tool, so to find significant grouping we iterate over a range of perplexities. We set the perplexity equal to 20 and the maximum iteration number to 1000. We also note that the definition of the groups found from this method are subjective; as we did this iterative process we found the global structure start to settle into groups that we defined visually. {Moreover, the standard usage of t-SNE does not handle data uncertainties well. The use of uncertainties in the algorithm hinders it ability to preserve the local structure of the data. We opt to not include uncertainties within our t-SNE analysis.} {In summation,} the classification of the populations within this low-dimensional map are subjective. In Section \ref{subsec: results - ML}, we investigate the distribution of these properties in connection to the coverage of AGN selections for each grouping.

%%%%%%%%%%%%%%%%%%%%%%%%%%%%%%%%%%%%%%%%%%%%%%%%%%%%%%%%%%%
%%
%%      Results
%%
%%%%%%%%%%%%%%%%%%%%%%%%%%%%%%%%%%%%%%%%%%%%%%%%%%%%%%%%%%%
\section{Results} \label{sec:results}
We present the results of the SED analysis here.
We have modeled the SEDs of dwarf galaxies with a variety of AGN signatures found in W24, with 461 SEDs models making our quality cut on $\chi_{r}^2 $. From these models, we collected Bayesian estimates for their host galaxy parameters. We draw connections between the subsamples and their host galaxy properties.

We used the t-SNE method to connect the array of selection techniques to the host galaxy parameters. t-SNE found nine new bins based on the similarity of the objects' host galaxy parameters. We then look at the properties for these new groups to investigate the influence of host galaxy further while trying to decouple their effects on galaxy selected by multiple methods. 

{The ADGD, including our sample, AGN selection results, and derived host galaxy and AGN parameters is available for public download on GitHub\footnote{https://github.com/erikwasleske/Active-Dwarf-Galaxy-Database}. } 

\subsection{AGN Selection subsamples}
\label{subsec: results- agn selected subsamples}
Here we investigate host galaxy properties for AGN selected via each technique in the ADGD.

\begin{figure}%
    \centering
    \includegraphics[width=0.49\textwidth]{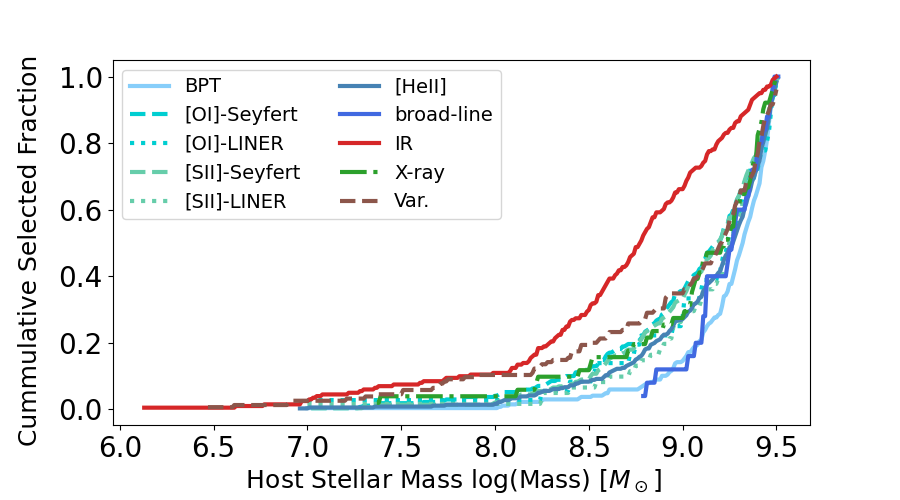} %
    \caption{ The cumulative fraction of objects selected as a function of host galaxy{'s} stellar mass for each sub-sample identified in W24.}
    \label{fig: subsample Mass funct}%
\end{figure}

We first investigated the {host galaxy stellar} mass distribution of our subsamples. Figure \ref{fig: subsample Mass funct} gives the cumulative sum of the selected fraction as a function of host galaxy stellar mass for each sub-sample. While most of the methods have comparable mass distributions, there are some interesting differences. 

\begin{figure}[h]
  \centering
    \subfigure{\includegraphics[width=0.475\textwidth]{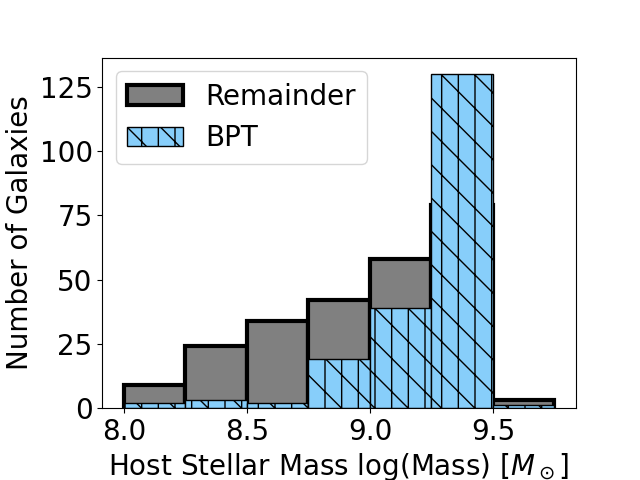}}%\quad
    \hspace{1em}% Space between image A and B
    \subfigure{\includegraphics[width=0.475\textwidth]{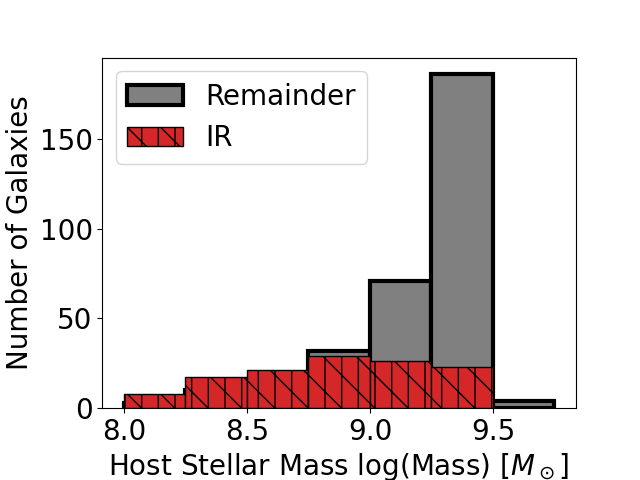}}
  \caption{\textit{Top:} Mass distribution of the BPT selected sub-sample  in comparison to the Mass distribution of the remaining objects in the database not selected by that method (grey). \textit{Bottom:} Same plot now separating our the IR selected sub-sample from the remaining database.  }
  \label{fig: BPT IR mass compare}
\end{figure}

IR color selected a significantly higher fraction of low-mass candidates compared to other sub-populations, with variability having the second most bottom-heavy mass function. 
Comparatively, the BPT and broad-line selections identify more massive candidates, suggesting these techniques might be biased towards AGN in more massive dwarf galaxies. These BPT and IR populations are compared to the remaining galaxies with the database in Figure \ref{fig: BPT IR mass compare}. These plots demonstrate the mass biases. Applying a two-sample Kolmogorov-Smirnov (K-S) test \citep{massey1951} to compare these sub-sample distributions to the remainder resulted in a near 0 p-values (much less than the stand threshold of $0.05$), pointing to the stark difference in these distributions.  We cautiously use this statistical test in comparing sub-sample host galaxy parameters as it relies on continuous data and is sensitive to outliers and extrema. We use it only as a reference guide to draw qualitative results.

The mass functions of the Seyfert and LINER populations, selected by either [\ion{O}{1}] or [\ion{S}{2}] emission, are in good agreement with each other. 

\begin{figure*}%
    \centering
    \subfigure{{\includegraphics[width=0.475\textwidth]{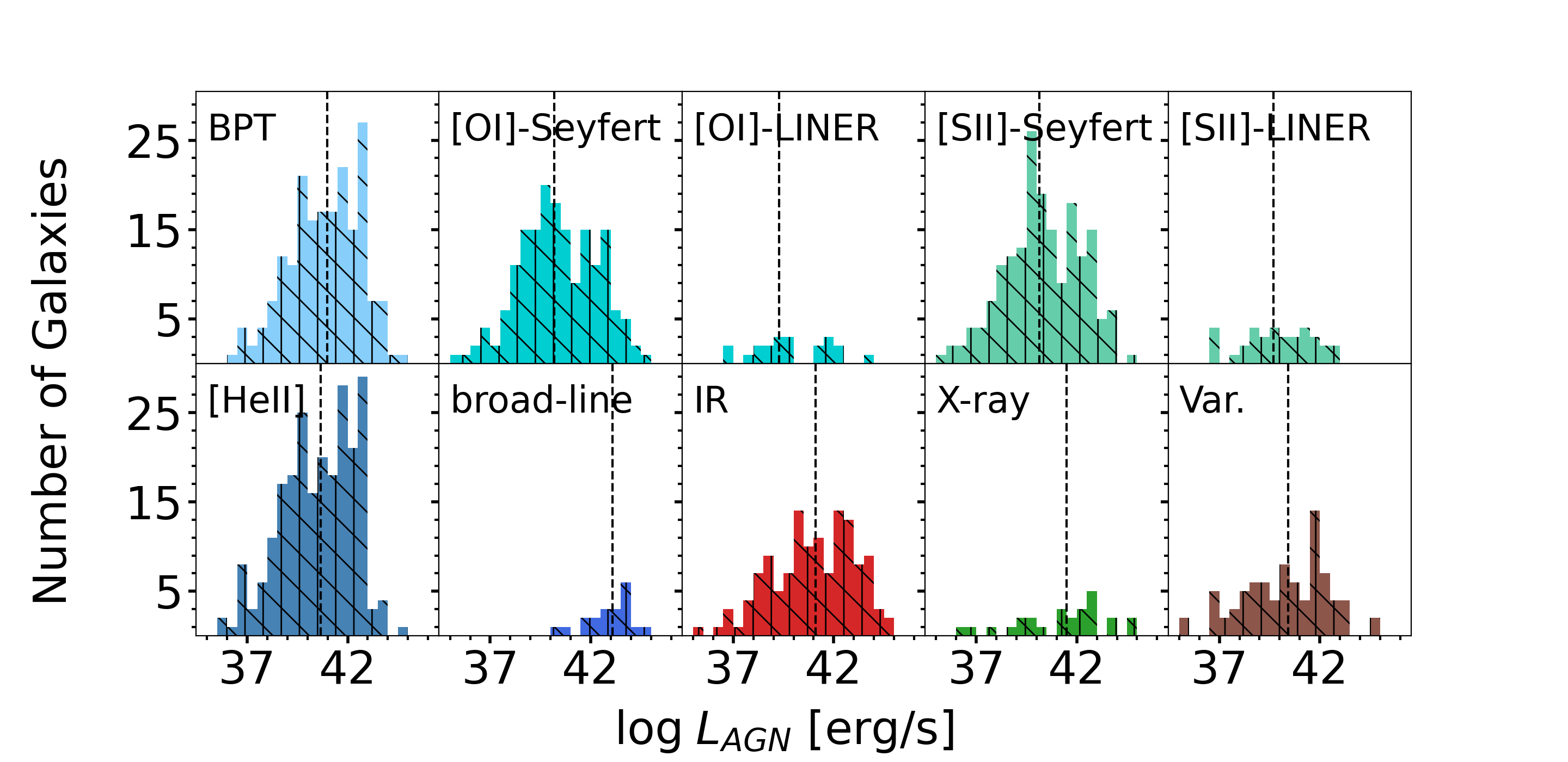} }}%
    \qquad
    \subfigure{{\includegraphics[width=0.475\textwidth]{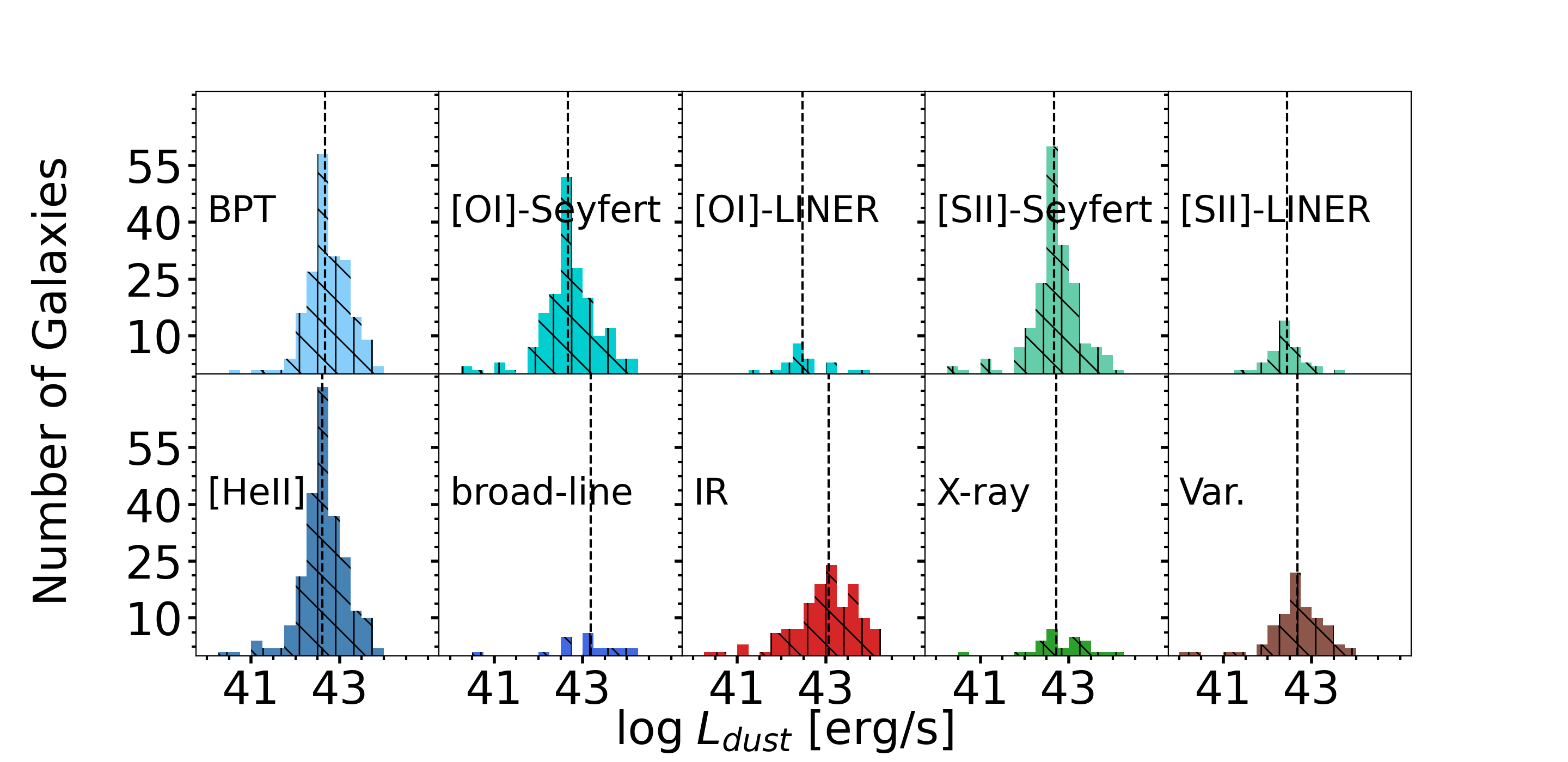} }}%
    
    \subfigure{{\includegraphics[width=0.475\textwidth]{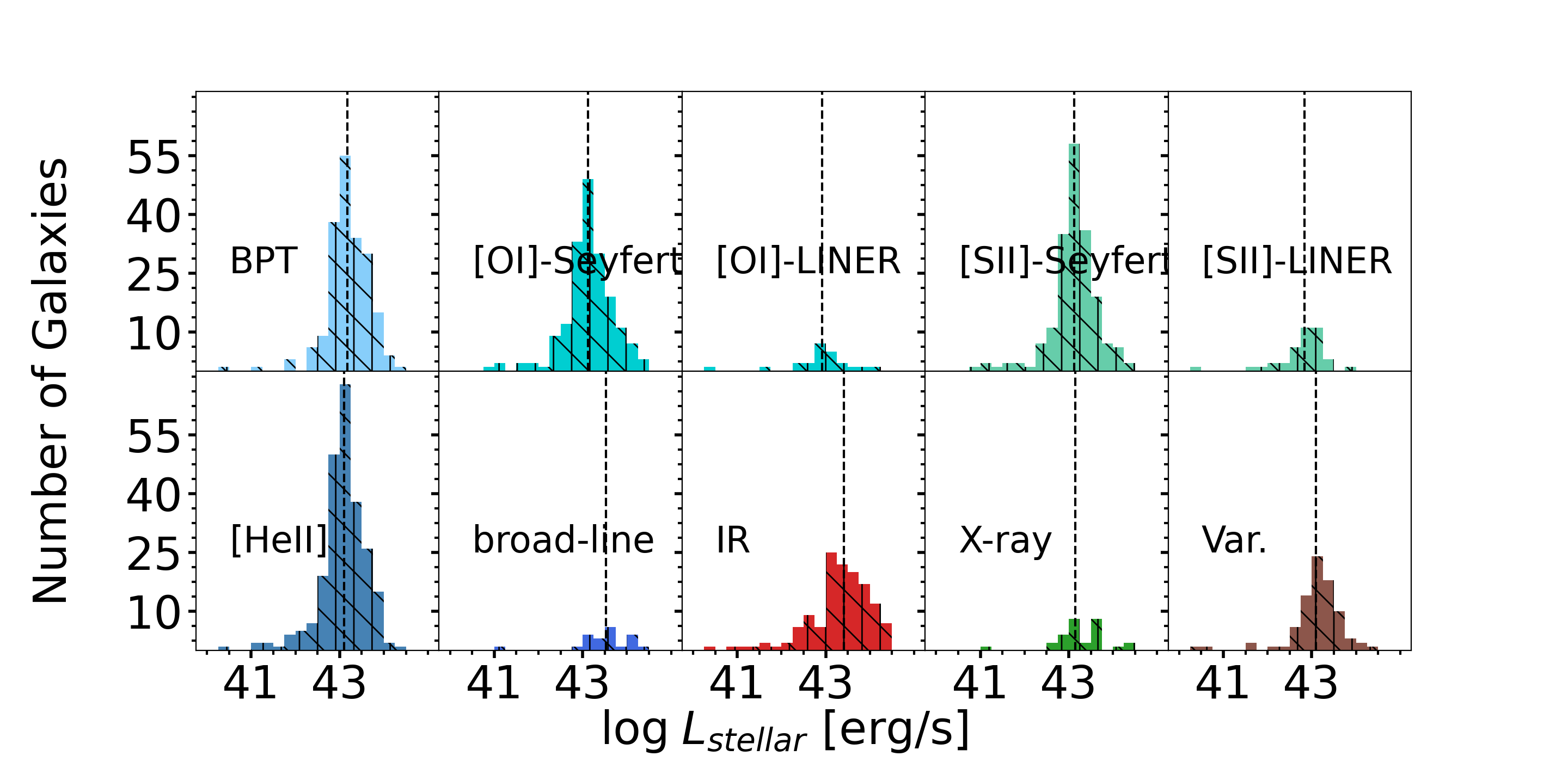} }}%
    \qquad
    \subfigure{{\includegraphics[width=0.475\textwidth]{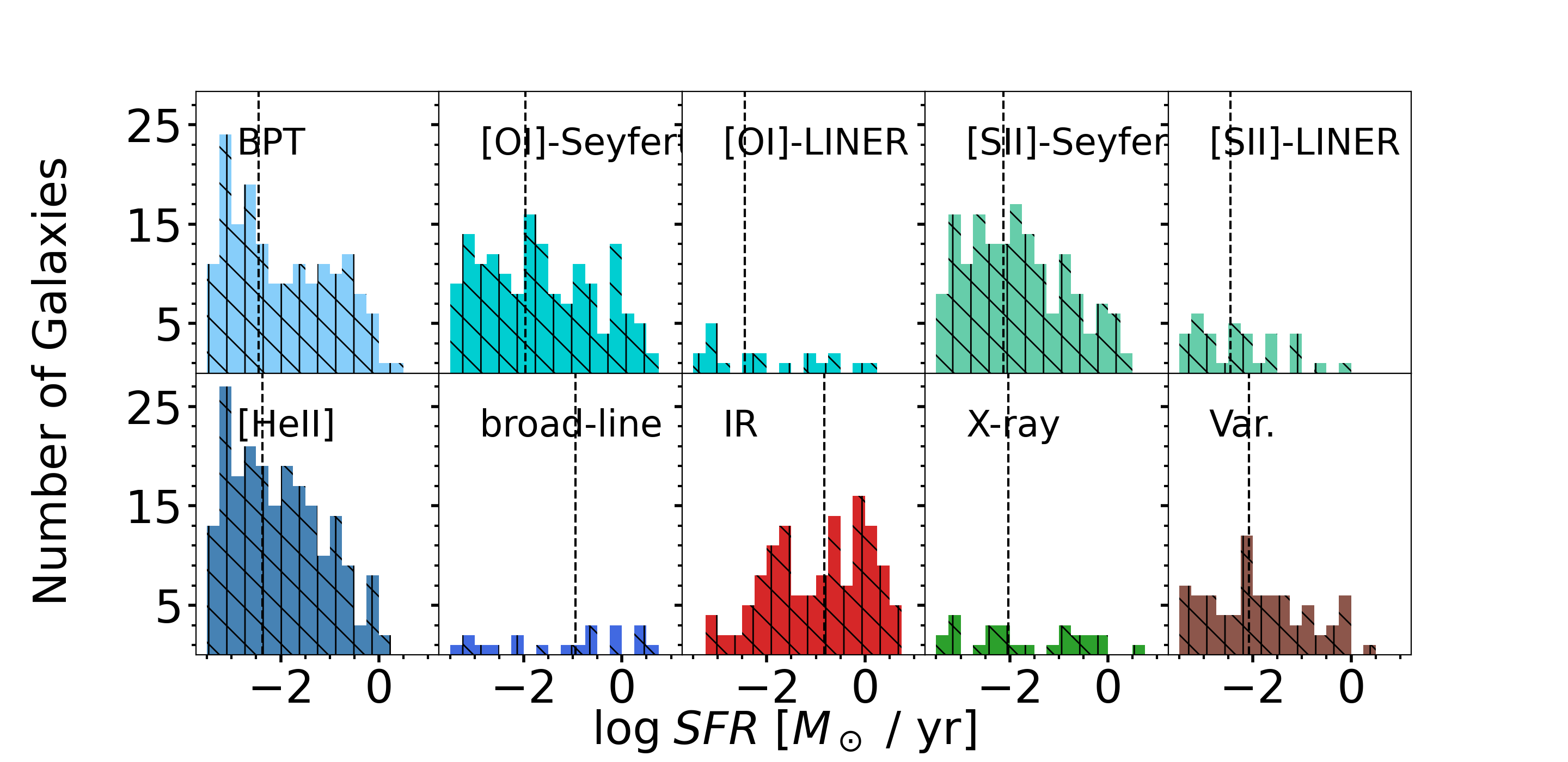} }}%
    \caption{
    Histograms of $L_{\text{AGN}}$, $L_{\text{dust}}$, $L_{\text{stellar}}$, and SFR estimates for subsamples of each AGN selection method. Median values of each distribution are given shown by the dashed line. The SFR histograms (\textit{bottom right}) is the logarithmic Bayesian estimated from \texttt{XCIGALE} for the SFR averaged over 100Myrs. {The mean uncertainties  for {$\text{log}\;L_{\text{AGN}}$, $\text{log}\;L_{\text{dust}}$, $\text{log}\;L_{\text{stellar}}$, and \text{log}\;SFR } are {0.85, 0.05 , 0.04, 0.31} respectively.}
    }%
    \label{fig: subsample parameter histograms}%
\end{figure*}

We also look at the distributions of host galaxy and AGN parameters in connection with the selection technique to investigate the potential biases of each method individually.
Table \ref{tab:subsample parameters} lists the {median} Bayesian estimated parameters from our SED models. {We additionally collect Bayesian estimates for the uncertainties of these values, which is the standard deviation of the distribution from all iterations of the model weighted by their probability \citep{boquien2019, yang2022}.} Figure \ref{fig: subsample parameter histograms} presents histograms of the $L_{\text{AGN}}$, $L_{\text{dust}}$, $L_{\text{stellar}}$, and SFR estimates for galaxies selected via each technique. {We quote mean Bayesian uncertainty for derived host galaxy parameters in Figures \ref{fig: subsample parameter histograms} and \ref{fig: f_agn histo}. The uncertainties are relatively small allowing us to draw conclusions about the differences in galaxy/AGN parameters between samples.}
%Recall that most AGN are selected via more than one technique, as such they can appear in multiple subsamples. 

\begin{figure}[h]
  \centering
      \subfigure{\includegraphics[width=0.475\textwidth]{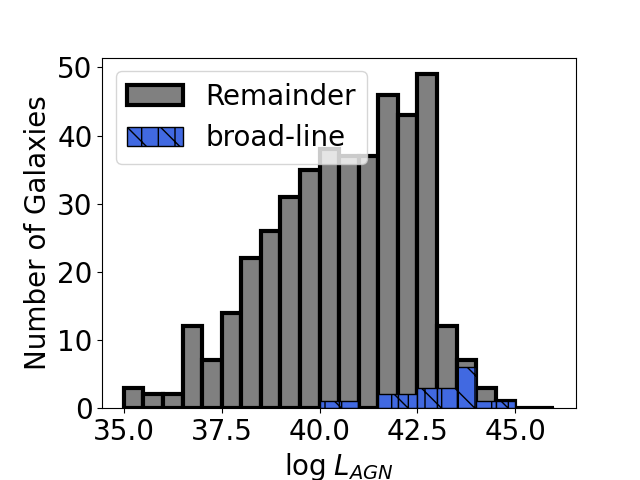}}%\quad
    \hspace{1em}% Space between image A and B
    \subfigure{\includegraphics[width=0.475\textwidth]{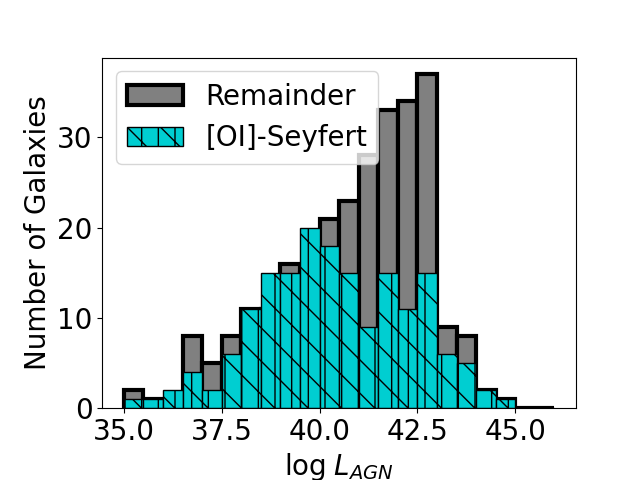}}%\quad
    \hspace{1em}% Space between image A and B
    \subfigure{\includegraphics[width=0.475\textwidth]{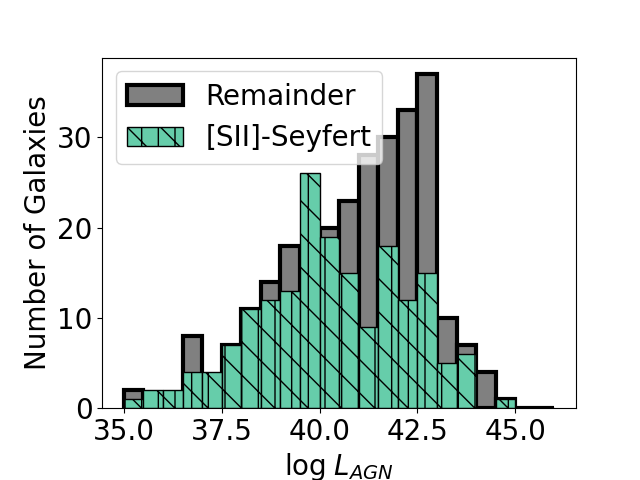}}
  \caption{ $L_{\text{AGN}}$ distribution of the broad-line selected sub-sample (\textit{top}),  [\ion{O}{1}] (\textit{middle}) and [\ion{S}{2}] (\textit{bottom}) Seyferts compared to the remaining galaxies in the database. 
  }
  \label{fig: Seyfert L_agn}
\end{figure}

\begin{figure}%
    \centering
    \includegraphics[width=0.50\textwidth]{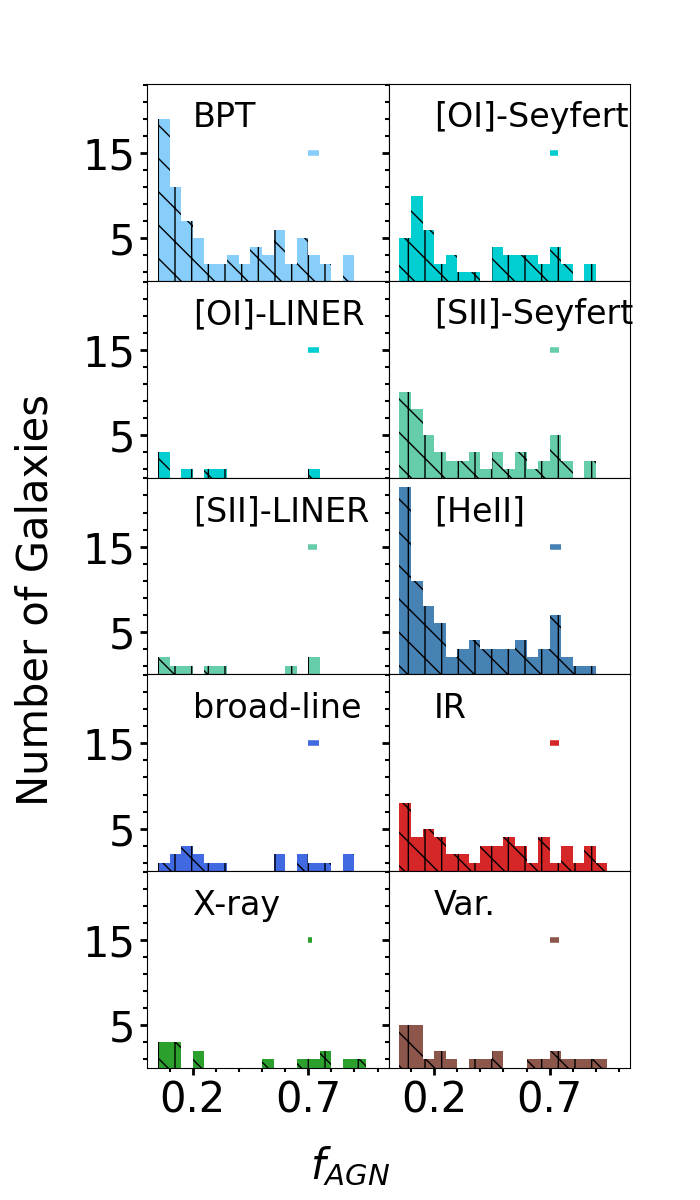} %
    \caption{{Histogram of fractional contribution of the AGN template to the IR luminosity,$\text{f}_{AGN}$, for each AGN selection method subsample. The color matched horizontal lines give the mean Bayesian estimate for the uncertainty for the values in each histogram, with mean error being 0.03.}}
    \label{fig: f_agn histo}%
\end{figure}

There are some differences in the $L_{\text{AGN}}$ distributions of the subsamples. AGN identified via broad-line emission comprise the smallest sample (though these have the highest overlap with the other subsamples; W24). We find the broad-line AGN to have the highest median $L_{\text{AGN}}$ by two orders of magnitude (median $L_{\text{AGN}} \approx 10^{43}\; \rm{erg\; s^{-1}}$) and the smallest spread of $L_{\text{AGN}}$.  Furthermore, the broad-line selected population as the highest median $L_{\text{dust}}$ and $L_{\text{stellar}}$. We show how the $L_{\text{AGN}}$ distribution compares to the rest of the database in the top panel of Figure \ref{fig: Seyfert L_agn} .
The comparison of the [\ion{O}{1}] and [\ion{S}{2}] Seyfert distributions are given in the bottom two panels of Figure \ref{fig: Seyfert L_agn}. These two-sample K-S test for these distributions are reporting near 0 p-value, with their shape pointing to a bias of less powerful AGN within these dwarfs compared to populations identified by the BPT, [\ion{He}{2}], and IR diagnostics. We also see that [\ion{O}{1}] and [\ion{S}{2}] LINERS have the least luminous AGN (median $L_{\text{AGN}} \approx 10^{39}\; \rm{erg\; s^{-1}}$). The remaining subsamples have median $L_{\text{AGN}}\approx 10^{40-41}\;\rm{erg\; s^{-1}}$.  

There is overall little variation in the dust and stellar luminosities between subsamples. In particular, populations selected by narrow optical emission lines techniques have nearly identical dust luminosity distributions. IR and broad-line selected samples had slightly higher median dust luminosities. Stellar luminosities are all relatively similar (median $L_{\text{stellar}}\approx10^{43}\;\rm{erg\;s^{-1}}$), though broad-line selected objects have slightly higher median stellar luminosities and LINERS have slightly lower. 

We do note some key differences in SFRs between the subsamples, though in general each technique has a wide spread in SFR. The BPT, He II, and LINER subsamples have the lowest median SFRs (median SFR $\approx 0.003\; \rm{M}_{\odot}/\rm{yr}$). 
These median SFR values are approximately half that of the Variability and X-ray subsamples.
Variability and X-ray selected AGN have higher medians; 0.006 and 0.007 $\rm{M}_{\odot}/\rm{yr}$, respectively. Broad-line AGN have median SFRs of $\sim0.1\; \rm{M}_{\odot}/\rm{yr}$, and IR selected AGN have the highest median, at $0.14\; \rm{M}_{\odot}/\rm{yr}$.

{In Figure \ref{fig: f_agn histo}, we present the histograms for the fractional contribution of the AGN IR luminosity to the total IR luminosity. This reflects the prevalence of the torus emission relative to the overall IR emission. The LINER, broad-line, X-ray and variability selected subsamples show a {bimodal distribution}. The broadline subsample has the most galaxies with high AGN fraction($\text{f}_{AGN} \geq 0.7$, ie. \citealt{carroll2021,wasleske2022}) with 11 of the 21 galaxies that meet this benchmark. 
}

\begin{figure*}%
    \centering
    \includegraphics[width=0.99\textwidth]{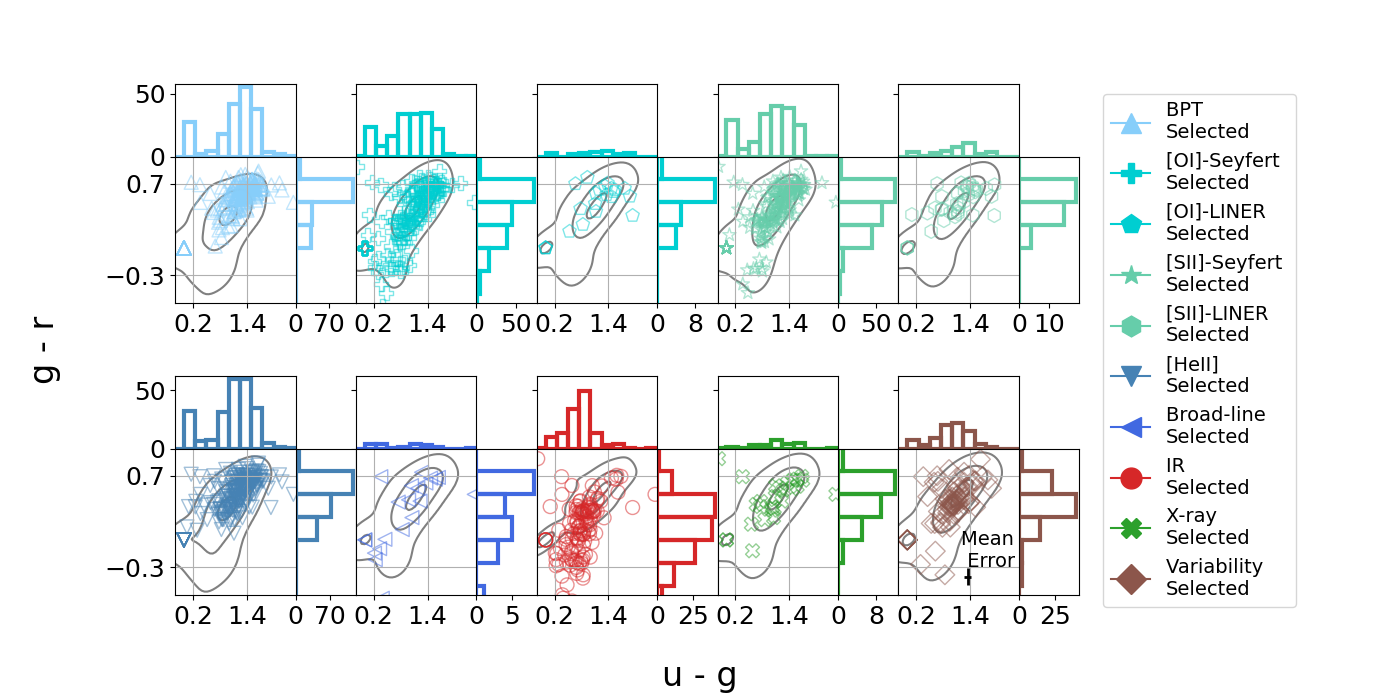} %
    \caption{SDSS u-g vs g-r diagram for the subsamples of AGN selected population shown on each panel. The grey contours show the 10, 50, and 90 percentile of the database \textit{not} selected by corresponded method for that panel (i.e. the Remainder of the database). The mean error of these color measurements for the full sample is shown in black in the last panel.}
    \label{fig: subsample optical color diagram}%
\end{figure*}

Next, we investigate the optical colors of our AGN subsamples. With the compiled photometry, we investigate the power-law slope of different aspects of the SEDs to discriminate between subsamples. 
Figure \ref{fig: subsample optical color diagram} shows the comparison of colors u-g and g-r using the de-reddened \texttt{cModelMag} from SDSS DR17.  
Immediately apparent is the blueness of IR selected AGN compared to their optical spectroscopic selected counterparts. In the u-g distributions, we see a separation between IR and spectroscopically selected population. The variability and X-ray selected AGN are more consistent with the other AGN populations. The IR selected population also has the biggest blue tail of its distribution for the g-r color. We can also see that the variability and X-ray g-r colors distributions are in between the IR and spectroscopically selected samples. The reddest objects are those selected with via the BPT and [\ion{He}{2}] diagrams. 
%These signatures point towards the active star-forming nature of those galaxies. 

\begin{figure*}%
    \centering
    \includegraphics[width=0.99\textwidth]{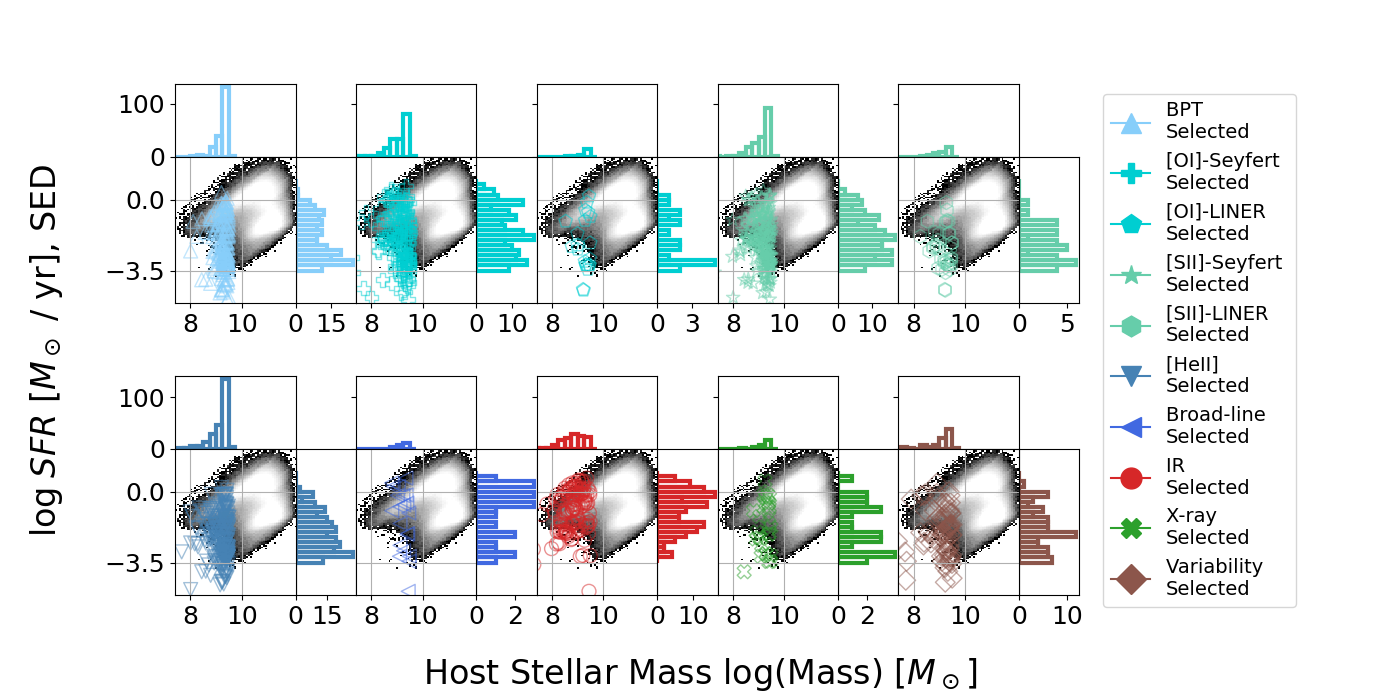} %
    \caption{
    We show the the SFR-Mass relation for the sub-populations of the 461 AGN in our analysis within each panel. In the background is a 2-D histogram of $\sim$640,000 galaxies within the GALEX-SDSS-WISE Legacy Catalog. We compare using our Bayesian estimates of average SFR over 100Myrs. The median uncertainty is $\sigma _\text{log SFR} = 0.09 \; [M_\odot/ yr]$.}
    \label{fig: subsample SFR-Mass}%
\end{figure*}

In Figure \ref{fig: subsample SFR-Mass}, we present the SFR - stellar mass relation for the AGN subsamples. We lay the AGN subsamples over a 2D histogram of $\sim$640,000 galaxies from GALEX-SDSS-WISE Legacy Catalog Shallow `All-sky' catalog (GSWLC-2, \citealt{salim2016,salim2018})\footnote{https://salims.pages.iu.edu/gswlc/}. The SFR and stellar masses of these galaxies are Bayesian estimates from SED modeling. These galaxies, with coverage across the UV-optical-IR, give a large population against which we can compare the physical properties of {AGN in} low-mass {galaxies}. Within the SFR-stellar mass space, we can clearly identify the blue cloud of star-forming galaxies and red sequence of quiescent galaxies \citep{daddi2007,rodighiero2011}. We note the wide distribution of the SFR for the active dwarf galaxies, with the IR selected population matching best to the low-mass regime of the blue cloud of star-forming galaxies. Those that are optically selected tend towards the red quiescent sequence of galaxies with lower SFR. 

A large fraction of the distribution of the optical, variability, and IR selected lay within the low-mass regime of the green valley, transitioning from eras of high star-formation to becoming `red and dead', whereas the X-ray selected population tends towards the bimodal distributions of the blue-cloud and red sequence. 

%%%
\subsection{t-SNE Clustering} \label{subsec: results - ML}
Below we discuss the results from the machine learning technique described in Section \ref{sec: Methods} to connect the host galaxy values to the diversity of AGN signatures found in W24.

\begin{figure*}
  \centering
    \includegraphics[width=0.742\textwidth]{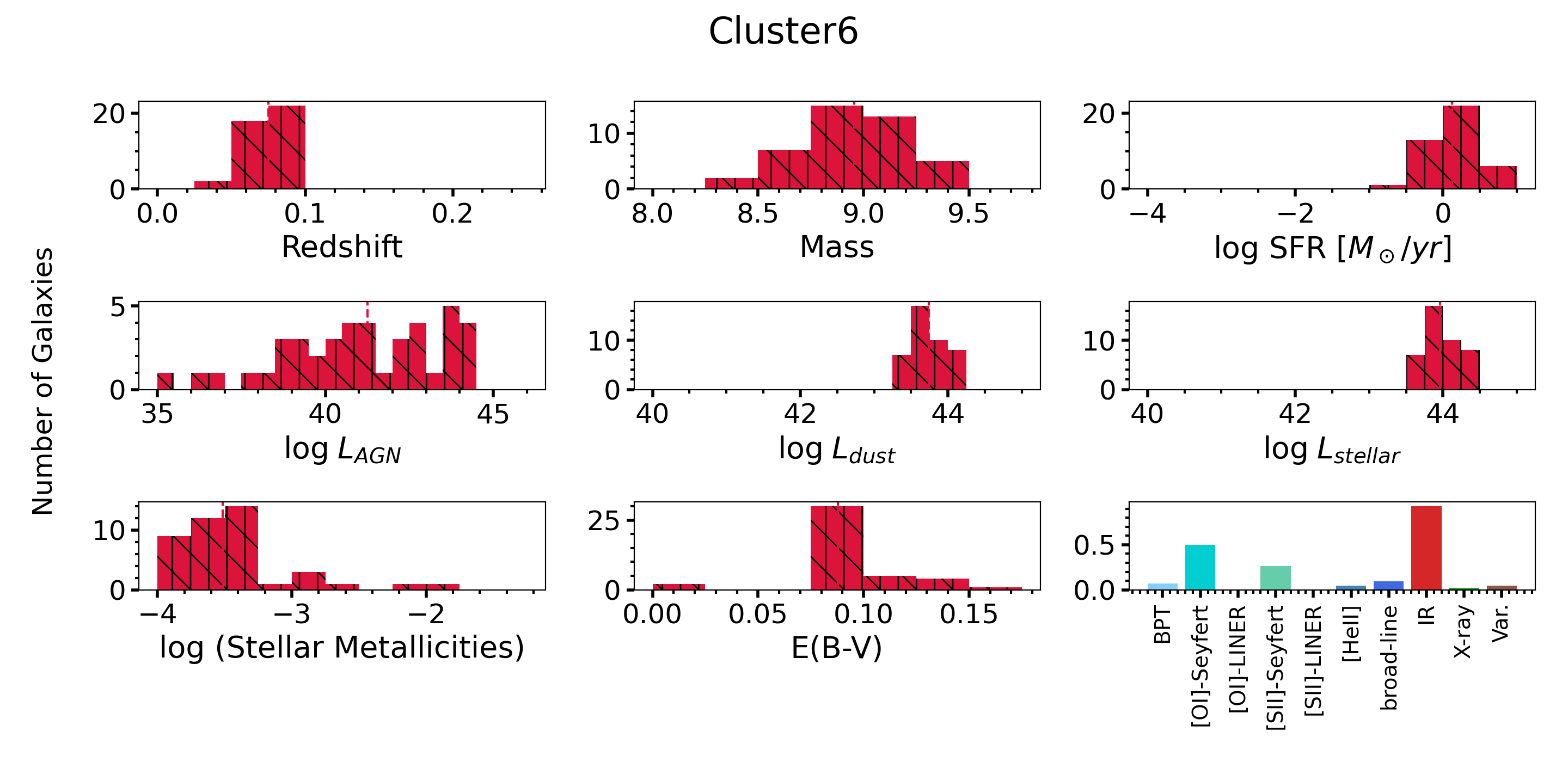}
  \caption{Distribution of host galaxy parameters for Cluster 6 from the t-SNE dimensionality reduction. 
  }
  \label{fig: tSNE cluster6 analysis}
\end{figure*}

From the implementation of t-SNE in Figure \ref{fig: tSNE clusters}, we separated the 2D map into the nine clusters shown. We analyze the distribution of the input parameters listed in Section \ref{subsec: methods - tSNE}. Table \ref{tab:cluster stats} gives the number of galaxies within each group and the median values for these parameters for each cluster. In Figure \ref{fig: tSNE cluster6 analysis} we show the distributions of these parameters within Cluster 6. The bottom right panel of the Figure shows the fractional value of each AGN selection technique within this cluster. Cluster 6 is dominated by those selected via IR colors with a fair overlap with Seyferts selected from the [\ion{O}{1}] and [\ion{S}{2}] diagrams. In the Appendix, Figures \ref{fig:cluster1-3 analysis}, \ref{fig:cluster4-5 7 analysis}, and \ref{fig:cluster8-9 analysis} show the distributions for the remaining clusters. 

\begin{figure}
  \centering
    \includegraphics[width=0.5\textwidth]{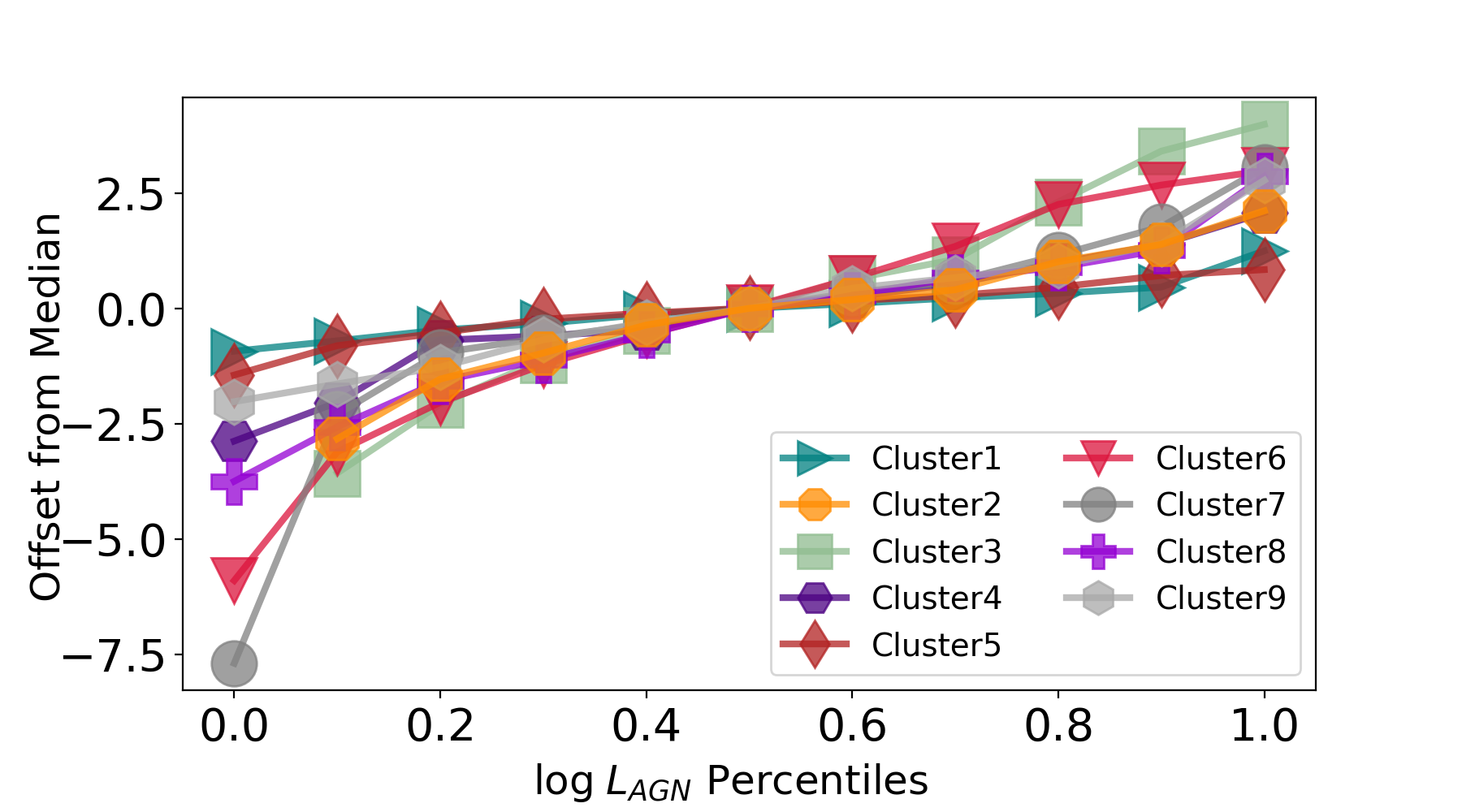}
  \caption{Percentile of the distribution vs. offset from median for each cluster within the t-SNE analysis for $L_{\text{AGN}}$. %The trends of Cluster 2 and 7 show how spread their $L_{\text{AGN}}$ distributions are. 
  }
  \label{fig: cluster percent Lagn}
\end{figure}

{In lieu of statistical values to determine the spread of the distribution of each parameter, we instead compare the shape of the distributions by evaluating the difference of the parameter value from the median value of its distribution for percentile steps within said distribution. We opt for this evaluation as distributions of the host galaxy parameters for each cluster deviate from normal distributions, with some having gaps within distribution. These plots thus give us a better understanding of the overall shape and dispersion of the values within }
%by evaluating the separation of the value of the parameter from the median for each cluster }
%We compare the shape of the distribution between clusters for each {of} the parameters using plots that give the offset from the median value at percentile values.
An example plot for $L_{\text{AGN}}$ is given in Figure \ref{fig: cluster percent Lagn}. {It shows the relation between the percentile steps through the distribution of $\text{log}(L_{\text{AGN}})$ for each cluster, compared to the difference of value in $\text{log}(L_{\text{AGN}})$ at that percentile to the median value of the distribution.}
Analogous figures for the remaining galaxy parameters are shown in the Appendix in Figures \ref{fig:Z, Mass percentile}, \ref{fig: SFR, dust percentile}, \ref{fig:Stellar, Metal percentile}, and \ref{fig: E(B-V) percentile}. Within these plots, heavier weights in distant bins from the median result in steeper slopes. Narrow distributions are represented by lines with slopes approaching zero. 

The t-SNE method was successful in finding groups dominated by optical narrow emission line selections (ie. Clusters 1, 2, 7, and 8) and those dominated by IR color selection (Clusters 4 and 6). Distinguishing significant contribution of X-ray and variability selected AGN within each cluster is limited by the smaller population of objects and thus AGN selection of galaxies within the ADGD. Those clusters with the highest variability selected fractions also have [\ion{He}{2}] selected fractions of $\sim 50\%$. We note the fraction of X-ray selected AGN scales with the fraction of BPT selected within each cluster.

 \begin{table*}[t]
    \centering
    \textbf{Median Cluster Host Parameter Values\\}
    \begin{tabular}{c | c c c c c c c c }
\hline
Cluster Number & Redshift & Mass & SFR & 
$\text{log}(L_{\text{AGN}})$ & $\text{log}(L_{\text{dust}})$ & $\text{log}(L_{\text{stellar}})$ & 
Metallicities & $E(B-V)$ 
\\

 &   &  [$\text{M}_\odot$] & [$\text{M}_\odot$ / yr] &
 [erg/s] & [erg/s] & [erg/s] &   &   \\

\hline
\hline

\begin{tabular}[c]{@{}l@{}}Cluster 1\\ Total: 46\end{tabular}& 0.0303 &9.30 & 0.0074 & 
42.44 & 42.57 & 43.08 &  
0.0016 & 0.0092
\\

\hline
\begin{tabular}[c]{@{}l@{}}Cluster 2\\ Total: 68\end{tabular}&0.0332&9.37&0.0089& 
41.43 & 42.57 & 43.09 &  
0.0075 & 0.0879
\\

\hline
\begin{tabular}[c]{@{}l@{}}Cluster 3\\ Total: 49\end{tabular}& 0.0246 & 8.4 & 0.0045 & 
38.82 & 42.34 & 42.70 & 
0.0004 & 0.0875
\\

\hline
\begin{tabular}[c]{@{}l@{}}Cluster 4\\ Total: 32\end{tabular}& 0.0568& 8.47 & 0.0841 & 
40.78 & 42.90 & 43.25 & 
0.0002 & 0.08772
\\

\hline
\begin{tabular}[c]{@{}l@{}}Cluster 5\\ Total: 20\end{tabular}& 0.0265 & 9.20 & 0.03820 & 
42.86 & 42.91 & 43.38 &
0.0027 & 0.1491
\\

\hline
\begin{tabular}[c]{@{}l@{}}Cluster 6\\ Total: 42\end{tabular}& 0.0752 & 8.96 & 1.3177 &
41.25 & 43.74  & 43.96 &   
0.0003 & 0.0878
\\

\hline
\begin{tabular}[c]{@{}l@{}}Cluster 7\\ Total: 97\end{tabular}& 0.0288 & 9.2 & 0.0014 &
39.20 & 42.55 & 43.015 &
0.0004 &  0.0875
 \\

\hline
\begin{tabular}[c]{@{}l@{}}Cluster 8\\ Total: 59\end{tabular}& 0.0439 & 9.4 & 0.0090 & 
40.41 & 42.96 & 43.42 &    
0.0003 & 0.0875
\\

\hline
\begin{tabular}[c]{@{}l@{}}Cluster 9\\ Total: 48\end{tabular}& 0.1059 & 9.28 & 0.1847 & 
42.03 & 43.26 & 43.58 &
0.0009 & 0.0877
\\

\hline

\end{tabular}
    \caption{ The median values of the host galaxy parameters estimated from the SED modeling for the clusters defined by the t-SNE dimensionality reduction.{The spread of these distributions are given by the percentile plots discussed in Section \ref{subsec: results - ML}}. }
    \label{tab:cluster stats}
\end{table*}

%%%%%%%%%%%%%%%%%%%%%%%%%%%%%%%%%%%%%%%%%%%%%%%%%%%%%%%%%%%
%%
%%      Discussion
%%
%%%%%%%%%%%%%%%%%%%%%%%%%%%%%%%%%%%%%%%%%%%%%%%%%%%%%%%%%%%
\section{Discussion} \label{sec:discussion}

We have estimated host galaxy and AGN properties for a set of active dwarf galaxies that were selected by a variety of electromagnetic accretion signatures. In evaluating the distribution and relations of these parameters for each sub-sample of dwarf AGN, we attempt to draw conclusions about the biases of each selection method.  
Then, we investigated the parameters distributions within the clusters of the t-SNE results to search for trends within the overlap of the selection methods. 
In the following sections, we discuss the implications of these results.

\subsection{Different evolutionary states of AGN subsamples}
\label{subsec: discussion - evolution differences}

Overall, the results we find from the distribution of host galaxy parameters for each sub-sample population provide quantitative support for previous qualitative understanding of the biases of each selection technique \citep{mendez2013,hainline2016, cann2019, burke2021, arcodia2024}. We find that our IR sub-sample contains high levels of star formation and is generally bluer in the optical. The optical spectroscopically selected AGN reside in the host galaxies with the lowest average SFRs in the sample. Our broad-line and X-ray samples contain relatively luminous AGN compared to the other subsamples. We can now discuss these results within the perspective of other studies.

Previous works have examined the selection of AGN in connection with the mass and age of the host galaxy, and the accretion process \citep{silverman2008, schawinski2009, hickox2009}. 
\cite{hickox2009} evaluated three different classes of non-dwarf, massive AGN, identified by X-ray, IR, and Radio emissions, at a redshift of $0.25 < \text{z} <0.8$. They found that the X-ray selected objects were in galaxies of all colors whereas the IR selected group comprised bluer and less luminous galaxies. The radio population was found to have very low Eddington ratios and the most massive BHs. The distribution parameters of these three classes points to an evolutionary scenario. They suggest that IR AGN are at an earlier phase of AGN-galaxy co-evolution, accreting at higher rates with smaller central BHs and higher SFR host galaxies than those selected by X-ray, with radio being the latest stage. Radio galaxies were found to have the most massive halos, with IR selected having the least massive.

\cite{ellison2016} investigated host galaxy SFRs of AGN selected optically, from mid-IR, and from radio emission, finding a high dependence of the SFR on how the AGN is selected, especially for obscured systems. This supports an evolutionary sequence of post-merger systems having a dependence on obscuration to increased SFRs. 
%Additionally, they find that optically selected AGN exhibit a large range of SFRs. 
Their results show that merger events serve as a catalyst for extreme SF and high accretion rates for absorbed AGN. They find enhanced SFR from the mid-IR selected AGN population and obscured galaxies in the X-ray. They conclude that IR selected AGN have the highest fraction of mergers within their study as identified in Galaxy Zoo. For an unobscured population, the results are less clear, as the un-obscured optically-selected AGN lack enhanced SFRs and are unlikely to be fueled by recent merger events. 

\begin{figure*}%
    \centering
    \includegraphics[width=0.99\textwidth]{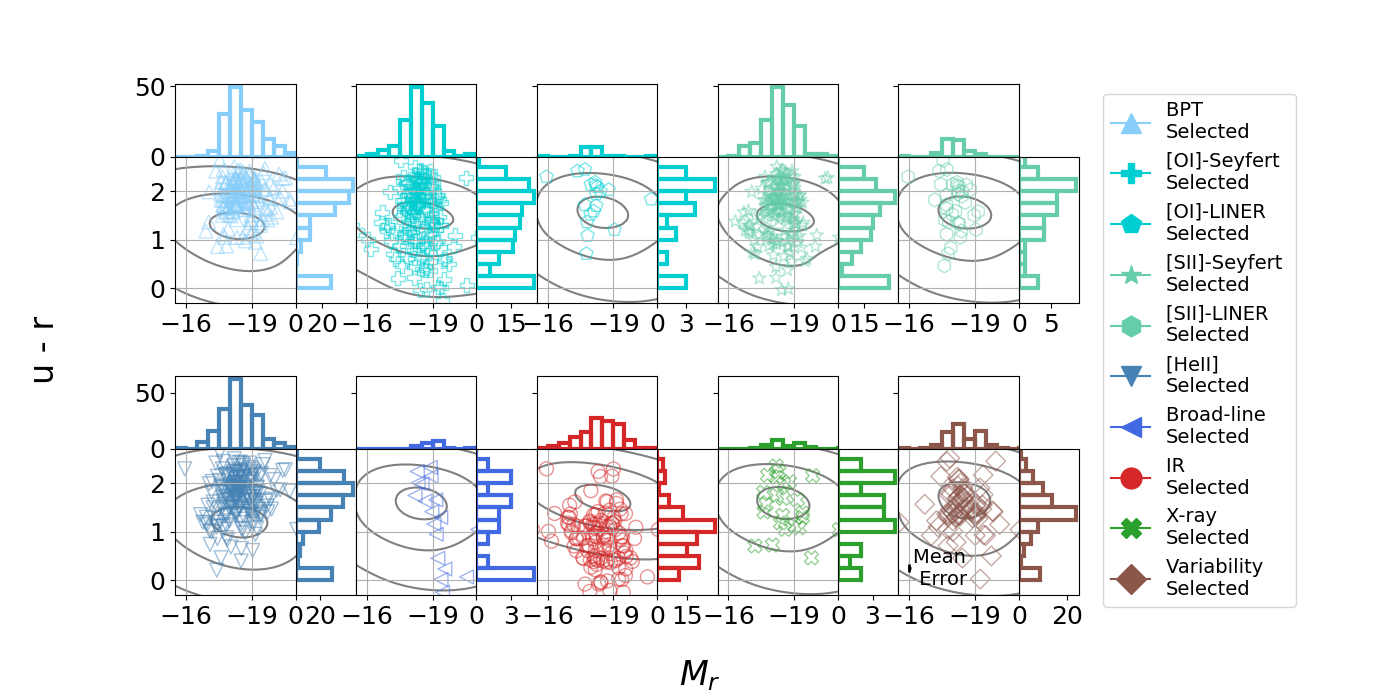} %
    \caption{
   Magnitude-color space of absolute magnitude in r band for each subsample, calculated from the \texttt{cModelMag} vs. the \texttt{cModelMag} u-r color. Grey contours and mean error bar are the same as in Figure \ref{fig: subsample optical color diagram}.
   }
    \label{fig: subsample M_r u-r}%
\end{figure*}

With our photometric data and estimated values of the host galaxy parameters, we can discuss our dwarf AGN subsamples within this evolutionary scheme. 
The star formation properties, stellar luminosity, and mass distribution suggests that the IR sub-sample is at an earlier evolutionary phase in comparison to our other populations. Figure \ref{fig: subsample M_r u-r} shows the magnitude color space of the subsamples. For similar distributions of r-band magnitude, there is evidence an evolution of IR-to-variability-to-optically selected with increasing u-r color, confirming the results of the SFR-Mass plot of Figure \ref{fig: subsample SFR-Mass}. We find greater separation between our X-ray and IR populations than the X-ray and IR selected populations of massive AGN with extended optical counterparts plotted on a similar diagram of Section 5 \cite{hickox2009}. {\cite{hickox2009} quotes their X-ray population as having mean magntiude and dispersion ($\langle\text{M}_r\rangle$, $\sigma_{\text{M}_r}$) values of (-21.3, 0.7) with mean color and dispersion ($\langle\text{u-r}\rangle$, $\sigma_{u-r}$) of (2.2, 0.4), and their IR population having values ($\langle\text{M}_r\rangle$, $\sigma_{\text{M}_r}$) = (-21.1, 0.6) and ($\langle\text{u-r}\rangle$, $\sigma_{u-r}$) = (2.1, 0.4).
Comparatively, our X-ray population has values of ($\langle\text{M}_r\rangle$, $\sigma_{\text{M}_r}$) = (-19.8, 4.6) and ($\langle\text{u-r}\rangle$, $\sigma_{u-r}$) = (1.5, 0.7). 
Our IR populations values are ($\langle\text{M}_r\rangle$, $\sigma_{\text{M}_r}$) = (-18.9, 3.9) and ($\langle\text{u-r}\rangle$, $\sigma_{u-r}$) = (0.8, 0.8). 
The distance between our X-ray and IR mean values within this color-magnitude space is 1.14, much greater than the separation of 0.22 between the median values of the massive populations from \cite{hickox2009}.
Overall our populations, being at lower redshift and mass, are less luminous and bluer. 
While our populations are more dispersed over this space, the trend of the massive X-ray populations being redder holds true and is more apparent than in the massive galaxy samples. 
This evidence supports the evolutionary model of AGN in that our X-ray subsample is an older redder phase, existing later in the green valley than the IR subsample. }

\subsection{{Are IR selected Dwarf AGN just Star Forming Galaxies?}} \label{subsec:dis IR sample}
\begin{figure}%
    \centering
    \includegraphics[width=0.49\textwidth]{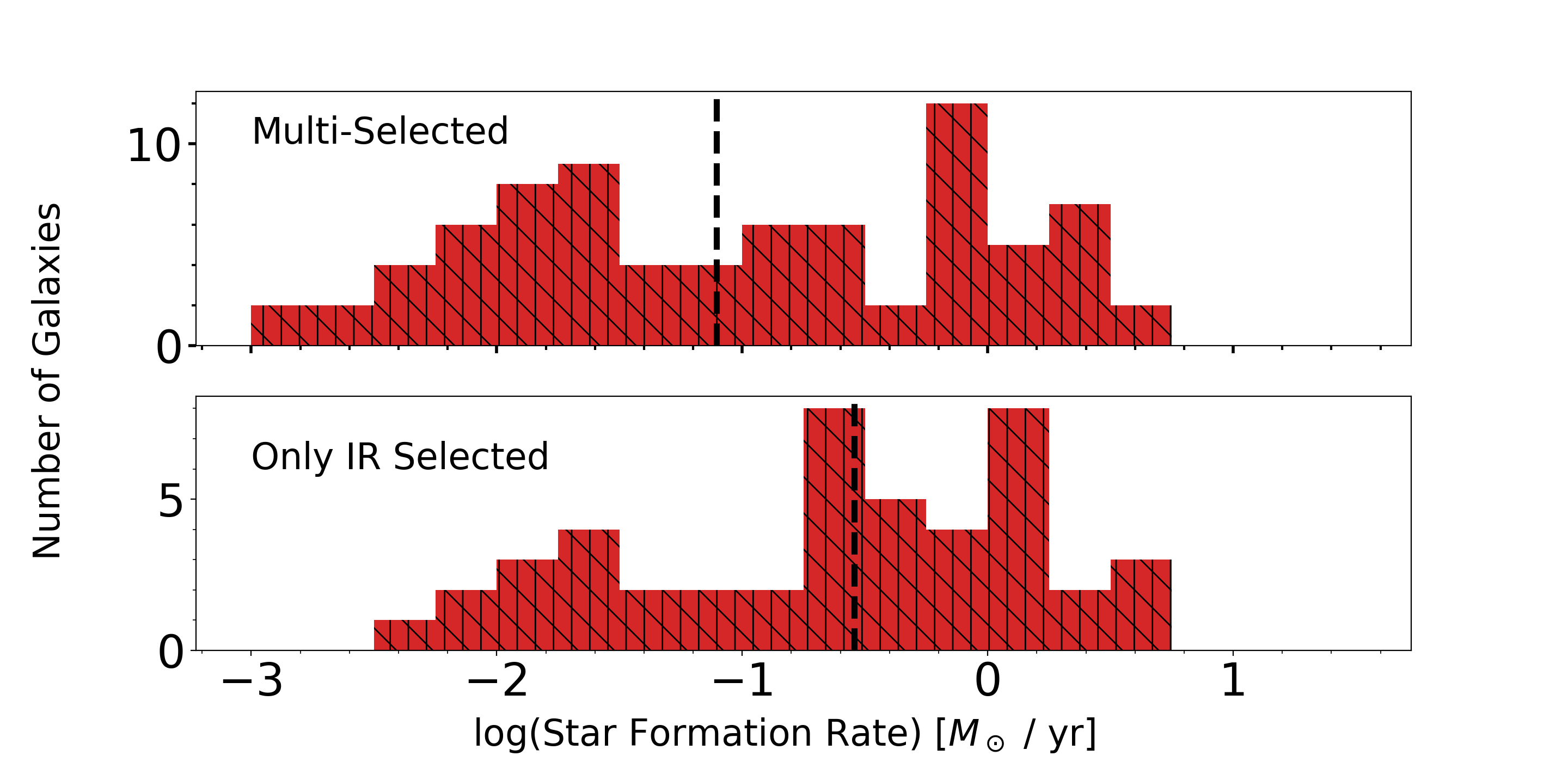} %
    \caption{ The histogram of SFR for the IR selected subsample of those selected by many method (\textit{top}) and those identified by their IR emissions (\textit{bottom}) alone. The vertical dashed lines give the median values of each distribution.}
    \label{fig: IR SFR histo separated}%
\end{figure}

{To investigate the concern of highly star-forming galaxies masquerading as AGN in IR color selections \citep{satyapal2018}, we seek to understand this sub-sample better.} As stated in W24, 457/702 of the AGN within the ADGD was selected by more than one method. Of the 201 dwarf galaxies selected as AGN from IR colors, 130 objects were selected by at least one additional method. Figure \ref{fig: IR SFR histo separated} breaks the SFR histogram of the IR selected sub-sample into those selected by another method from those only selected from IR colors. {Of the 71 objects selected by IR colors only, 47 have non-zero fractional contribution of the AGN template to their SED. However, the median AGN fraction of these 47 objects is 0.02, with the maximum value being 0.66 in this subset. The remaining 24 objects do not have SED modeling that met our criteria $ \chi_{r}^2$ mentioned in Section \ref{subsec: methods - SED}. As mentioned above, these objects were not included in our analysis. The small non-zero AGN SED contribution for the remaining 47 IR-only selected objects adds to ambiguity surrounding the purity of the IR selection.}
As the remaining IR-only group contains very few objects with low SFRs, this points to this method's bias of identifying high star-forming dwarf galaxies. 
{Additionally, t}he contribution of dust within the IR selected sub-sample in unison with the stellar emission and SFR shows the caution needed when implementing IR selection as young star-forming dwarfs can mimic AGN {colors} \citep{hainline2016}.

%With our optically selected subsamples having distributions of mid- to low-SFR {there} is supporting evidence of this evolutionary model of AGN selection within the dwarf galaxy mass regime. The overlap between our subsamples muddy the lines of this evolution. Follow-up morphology studies using high-resolution imaging could inform us of the merger history of these subsamples. 

%% 
\subsection{AGN Signatures and Host Parameter Distributions of the t-SNE Groups}
\label{subsec: discussion - t-SNE groupings}

The re-binning of groups based on AGN selection techniques (W24) to similarities in host galaxy parameters via t-SNE opens a new avenue to examine the disagreement between AGN selection methods. This dimensionality reduction method allowed us to explore the intersections of the Venn diagram of the AGN selection techniques applied to the ADGD.

The two most distinct groupings in this low-dimensional map are Clusters 1 and 2. These two assemblies have the most spread distributions of $L_{\text{dust}}$ and $L_{\text{stellar}}$. Both groups contain galaxies selected via a variety of methods, with the most prominent technique being [\ion{He}{2}] selection. Cluster 1 also has distinctively low E(B-V) compared to the high E(B-V) of Cluster 2. This points to [\ion{He}{2}] being less sensitive to the influences of dust and stellar emission in active dwarfs. 

We find that for Clusters 1, 2, 7, and 8 which have the highest fraction of spectroscopically selected objects all have similar medians and distributions of $L_{\text{stellar}}$. Clusters 2, 7 and 8 have lower median $L_{\text{AGN}}$ and wider distributions of $L_{\text{AGN}}$ than Cluster 1 (Figure \ref{fig: cluster percent Lagn}). 
Of these groups, Cluster 1 is the most purely [\ion{He}{2}] selected while having the {third} highest stellar metallicities. This is consistent with the [\ion{He}{2}] being more sensitive to the hardness of the ionizing source \citep{sartori2015}.

Clusters 4 and 6 have a significant fraction of IR selected ($>65\%$) objects. The second and third most significant selection method within both of these groups are Seyferts from the [\ion{O}{1}] and [\ion{S}{2}] diagrams. The most apparent differences between these two clusters is the wide distribution of SFR and $L_{\text{AGN}}$ for Cluster 6. Comparing to the other seven groups, Cluster 4's separation is dominated by the low mass of its objects.

Clusters 3, 5, and 9 are the groupings with the most diverse contributions of AGN selection methods. Cluster 3 lacks the contribution from BPT selection that Clusters 5 and 9 have, while having the highest contribution from variability selected of the three groups. 
Cluster 5 has significant selection fraction as BPT, Seyfert, and IR. This population has the highest obscuration among the clusters.

{In summary, the binning from t-SNE can be used to investigate the overlap of AGN selection techniques.
For the spectroscopically dominated clusters (1,2,7,8) compared the remaining groups, there is an inherent bias of spectroscopic selection towards selecting low SF, higher mass candidates than those selected via photometric methods. No cluster corresponds to a single selection method. The assortment of sub-sets within each group can provide insight into the scatter within the parameter spaces utilized by each selection technique. The lack of clean division of selection subsamples within the t-SNE clusters reflects the inherent degeneracies in computing host galaxy parameters and the difficulty of disentangling different selection effects.
}

{For example, in looking at Figure 7 of W24, we see that variability selected subsample is scattered across the [\ion{He}{2}] diagram, with a fraction of the subset lacking strong [\ion{He}{2}] and [\ion{N}{2}] emission. Within our t-SNE analysis, Cluster 3 has strong contributions of the variability and [\ion{He}{2}] samples within its populations, pointing to a possible underlying cause for this separation in the variable population on the [\ion{He}{2}] diagram. Cluster 3 represents a population that have less luminous AGN, and are forming less stars compared to the variability sub-sample as a whole. However, the diverse contributions of the selection subsamples within each Cluster contaminate the ability to infer a connection, as Cluster 3 also has a high IR selected population, shifting the stellar and dust luminosity distributions higher and the mass distribution lower. 
}

{Comparatively, Cluster 8 contains heavy contributions from Variability and [\ion{He}{2}] but lacks an IR population in favor of higher BPT contribution. From this, we see a brighter AGN luminosity, even lower SFR and a more massive sample than in Cluster 3. 
Looking at the [\ion{He}{2}] diagram of Figure 8 in W24, the scatter maybe do to the AGN luminosity and SFR, as the BPT selected population occupies a region of higher [\ion{He}{2}] and [\ion{N}{2}] emission compared to those Variability objects that are selected on this diagram. To get a full understanding of this interplay, a full analysis of the location of each object within each parameter space matched to the t-SNE cluster would be required. The degeneracy and contamination of all selected sub-samples contributing to all t-SNE clusters leads to the need to view of this problem on an object-to-object basis. }

\subsection{Need for New SED Templates for Dwarf AGN}
\label{sec: discussion -  new SED template}

Over the course of this work, we analyzed the results of host galaxy parameters estimated from SED models. However, it is important to recall that these templates were constructed for more massive host galaxies and their active nuclei. The results presented are dependent on the accuracy of the decomposition of these SED models into the intrinsic emission components in the galaxy. The difference between an accurate dwarf AGN template and the massive AGN templates used is unclear. %This uncertainty leads to unsure decomposition of the model templates, resulting in wrongly attributing emissions to stellar, dust or the AGN. 

The construction of a composite dwarf AGN SED template requires high{-}resolution photometry of active dwarf galaxies. We have discussed the difference in the SED for our various subsamples in this study, thus a focus on unobscured, bright dwarf AGN would provide the most ideal population to build this initial dwarf AGN template.  

%% 
%%%%%%%%%%%%%%%%%%%%%%%%%%%%%%%%%%%%%%%%%%%%%%%%%%%%%%%%%%%
%%
%%      Conclusions
%%
%%%%%%%%%%%%%%%%%%%%%%%%%%%%%%%%%%%%%%%%%%%%%%%%%%%%%%%%%%%
\section{Conclusions} \label{sec:conclusions}

Using the database of active dwarf galaxies compiled in W24, we investigated the connection between their accretion signatures and host galaxy properties.  Of the 702 AGN within the ADGD, we develop 461 statistically significant SED models from which to estimate host galaxy parameters. We then attempted to re-bin objects based on these properties. We gathered additional photometric data to construct SED models using \texttt{XCIGALE} \citep{yang2022}. Our models provided Bayesian estimates of the host galaxy properties. These values were used to draw connections to the AGN selection subsamples as well as serving as input for t-SNE to take the high-dimensional parameter space of redshift, {host galaxy stellar} mass, SFR, stellar metallicity, $L_{\text{dust}}$, $L_{\text{AGN}}$,$L_{\text{stellar}}$ and E(B-V) color to develop a low dimensional representation of the ADGD. 
Our conclusions are as follows:

\begin{itemize}
    %\item Of the 702 AGN within the ADGD, we develop 461 statistically significant SED models from which to estimate host galaxy parameters.

    \item The cumulative mass function shows IR and Variability subsamples biased towards {AGN in} less massive {galaxies}, whereas the BPT and broad-line selections are biased towards more massive active dwarf galaxies.

    \item The broad-line subsample has the brightest median $L_{\text{AGN}}$, while LINER objects have the lowest median $L_{\text{AGN}}$. The [\ion{O}{1}] and [\ion{S}{2}] Seyfert populations have a distinctly different $L_{\text{AGN}}$ distributions form the rest of the database.

    \item The $L_{\text{dust}}$ distributions are similar in median values and variations across subsamples. 

    \item The BPT, [\ion{He}{2}], and LINER subsamples have a median SFR that is half of the median value for the Variability and X-ray subsamples. 

    \item Our results within magnitude-color space suggest agreement with an evolutionary picture of our subsamples. This defines our IR sample as being in an earlier phase and the optically selected objects representing the later stages of this evolution. The X-ray and variability populations serve as the middle ground within this interpretation. 

    \item Future studies into the morphologies of these systems can be used to investigate the merger histories of these populations to search for further evidence of this evolution. 

    \item t-SNE provided a visualization of the ADGD within a low-dimensional parameter space to try to find connections between objects selected as AGN from multiple methods. However, the degeneracies of these populations does not lead to a fulfilling conclusions outside of re-evaluation of the host parameters in connection to a new binning schematic.

    \item {In evaluating our results for the most complete suite of AGN selection methods, we note that if only  [\ion{He}{2}] and IR selections are applied, 73\% of our database would be recovered, the highest rate of recovery for any combination of two methods. Throughout our analysis, we noted the stark differences between populations selected by these methods, along with t-SNE clustering emphasizing the difference between the populations selected by spectroscopy and IR. Furthermore, the combination of  [\ion{He}{2}], IR and Variability finds 88.3\%  of the total database, the highest recovery rate of any suite of three selection methods used.
    }

\end{itemize}

Overall, we find evidence for evolutionary differences between the subsamples of our dwarf galaxy database. The degeneracies of these populations starts to muddy these results. However, we note that this work is highly dependent upon the accuracy of our SED models and its decomposition. Currently, there exists no empirical model of a broadband SED for active dwarf galaxies. 

Decomposition of high resolution imaging of active dwarf galaxies across the electromagnetic spectrum can help to disentangle these issues and provided a more accurate prescription for the SED models of dwarf galaxies.

%%%%%%%%%%%%%%%%%%%%%%%%%%%%%%%%%%%%%%%%%%%%%%%%%%%%%%%%%%%
%%
%%      Acknowledgments
%%
%%%%%%%%%%%%%%%%%%%%%%%%%%%%%%%%%%%%%%%%%%%%%%%%%%%%%%%%%%%
%\section{Acknowledgments} \label{sec:acknowledgments}
\begin{acknowledgments}

The underlying data set used in this work and \cite{wasleske2024} is available in Zenodo at \dataset[doi: 10.5281/zenodo.17407817]{https://urldefense.com/v3/__https://doi.org/10.5281/zenodo.17407817__;!!DZ3fjg!-XoKkS0TPc-1jd_XVpitsceyWr-tM98V-1gko_cURhr4xNqXXqZerbwZ4lzXr3fg8iVOyDVYrly2RQ6_4s8Rtco$ }. The catalog is in a FITS format and contains positions, redshifts, photometry, emission line luminosities, SED fitting parameters, and AGN selection indicators.

This work is supported by the NASA Astrophysics Database Analysis Program through Grant 80NSSC23K0460.  

We thank the anonymous referee for comments that have improved this manuscript. 

Funding for the SDSS and SDSS-II has been provided by the Alfred P. Sloan Foundation, the Participating Institutions, the National Science Foundation, the U.S. Department of Energy, the National Aeronautics and Space Administration, the Japanese Monbukagakusho, the Max Planck Society, and the Higher Education Funding Council for England. The SDSS Web Site is http://www.sdss.org/.

The SDSS is managed by the Astrophysical Research Consortium for the Participating Institutions. The Participating Institutions are the American Museum of Natural History, Astrophysical Institute Potsdam, University of Basel, University of Cambridge, Case Western Reserve University, University of Chicago, Drexel University, Fermilab, the Institute for Advanced Study, the Japan Participation Group, Johns Hopkins University, the Joint Institute for Nuclear Astrophysics, the Kavli Institute for Particle Astrophysics and Cosmology, the Korean Scientist Group, the Chinese Academy of Sciences (LAMOST), Los Alamos National Laboratory, the Max-Planck-Institute for Astronomy (MPIA), the Max-Planck-Institute for Astrophysics (MPA), New Mexico State University, Ohio State University, University of Pittsburgh, University of Portsmouth, Princeton University, the United States Naval Observatory, and the University of Washington.

This publication makes use of data products from the Wide-field Infrared Survey Explorer, which is a joint project of the University of California, Los Angeles, and the Jet Propulsion Laboratory/California Institute of Technology, funded by the National Aeronautics and Space Administration.

This research has made use of data obtained from the Chandra Data Archive and the Chandra Source Catalog, and software provided by the Chandra X-ray Center (CXC) in the application packages CIAO and Sherpa

This research has made use of data obtained from the 4XMM XMM-Newton serendipitous source catalog compiled by the 10 institutes of the XMM-Newton Survey Science Centre selected by ESA.
\end{acknowledgments}

\software{Astropy \citep{astropy2018}, \texttt{astroquery} \citep{astroquery2019}, Matplotlib \citep{matplotlib2007}, NumPy \citep{numpy2020}, Scikit-learn \citep{scikit-learn}.
}

%%%%%%%%%%%%%%%%%%%%%%%%%%%%%%%%%%%%%%%%%%%%%%%%%%%%%%%%%%%
%%
%%      Bibliography
%%
%%%%%%%%%%%%%%%%%%%%%%%%%%%%%%%%%%%%%%%%%%%%%%%%%%%%%%%%%%%
\bibliography{ewasleske}{}
\bibliographystyle{aasjournal}

%%%%%%%%%%%%%%%%%%%%%%%%%%%%%%%%%%%%%%%%%%%%%%%%%%%%%%%%%%%
%%
%%      Appendix
%%
%%%%%%%%%%%%%%%%%%%%%%%%%%%%%%%%%%%%%%%%%%%%%%%%%%%%%%%%%%%
\appendix
%\restartappendixnumbering

\section{Parameter Distributions of Clusters}
Below are shown the parameter distributions for the other eight clusters mentioned in Section \ref{subsec: methods - tSNE}.

\begin{figure*}[h]
    \centering
    \subfigure{\includegraphics[width=0.742\textwidth]{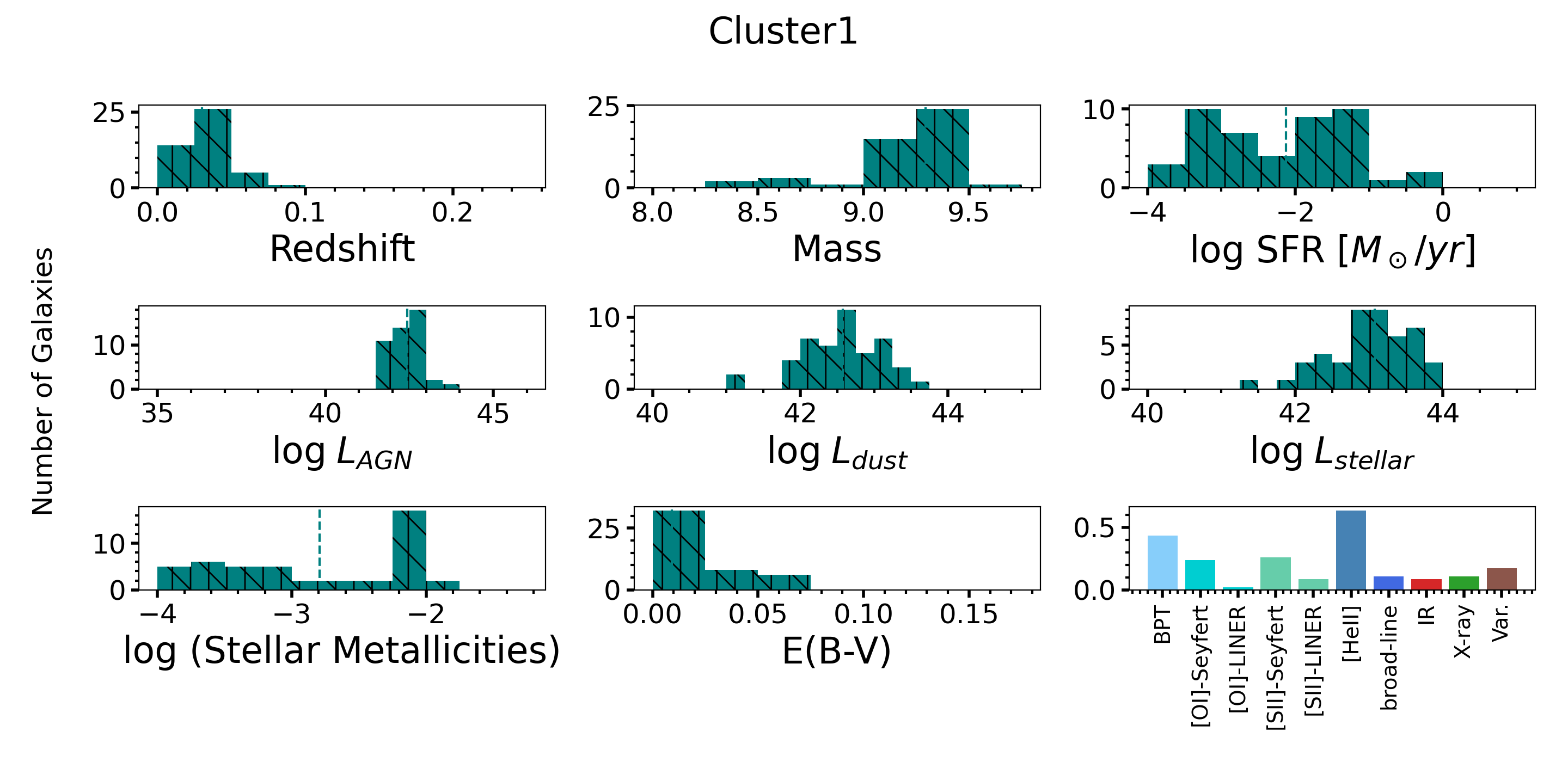} }
    \subfigure{\includegraphics[width=0.742\textwidth]{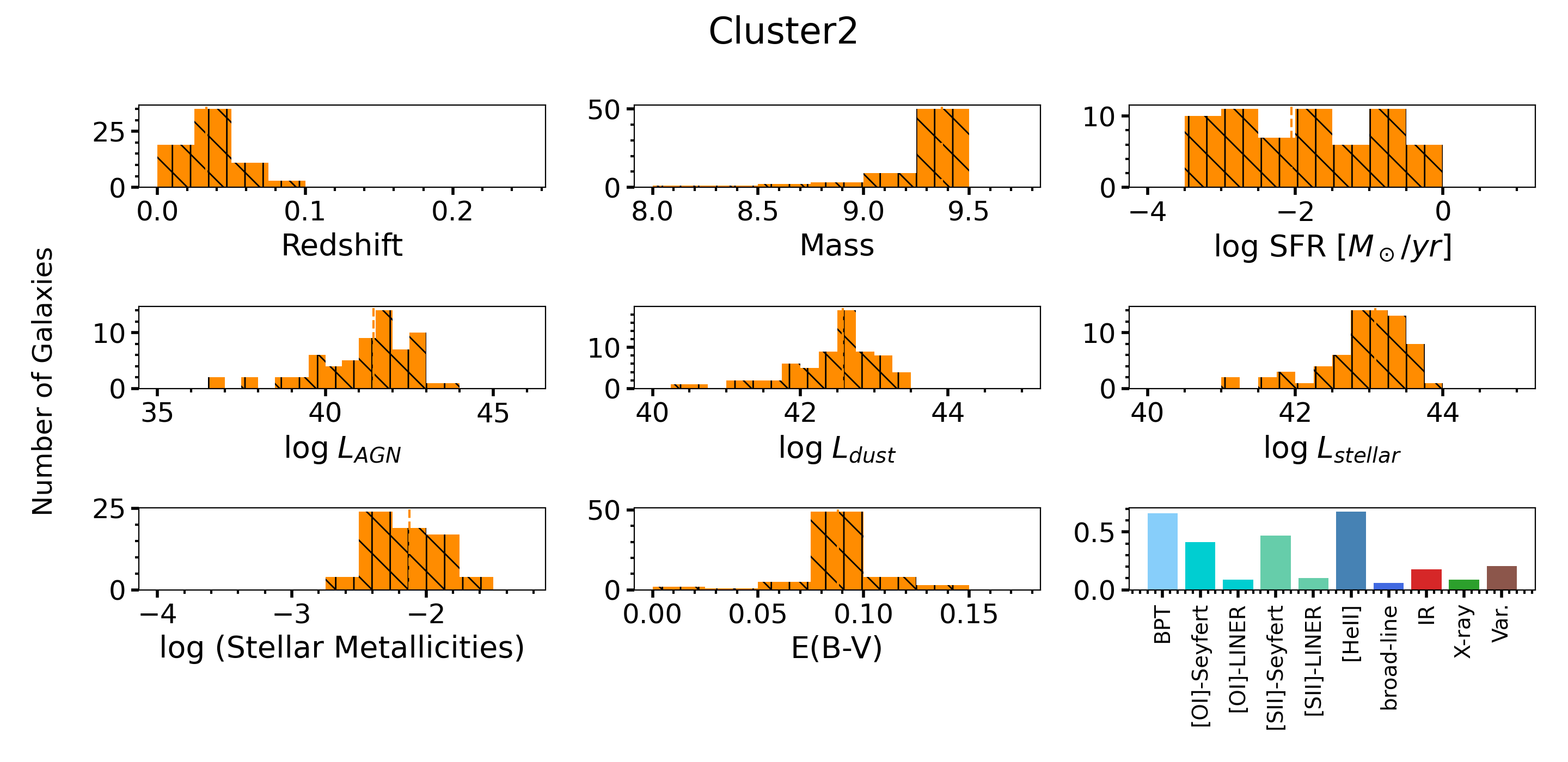} }
    \subfigure{\includegraphics[width=0.742\textwidth]{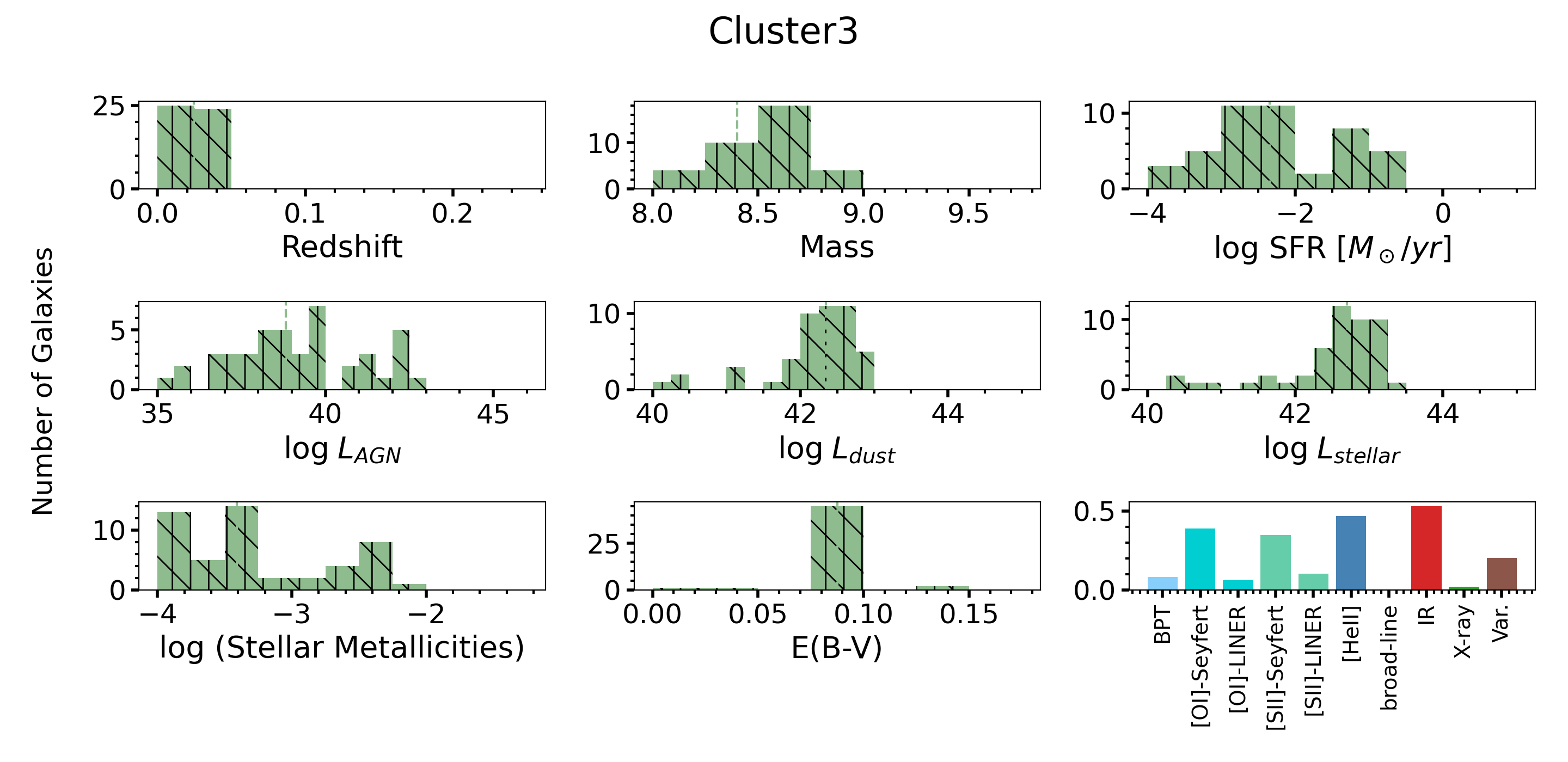} }
        \caption{ Distribution of host galaxy parameters for Clusters 1, 2, and 3 from the t-SNE dimensionality reduction.}
    \label{fig:cluster1-3 analysis}
\end{figure*}

\begin{figure*}[h]
    \centering
    \subfigure{\includegraphics[width=0.742\textwidth]{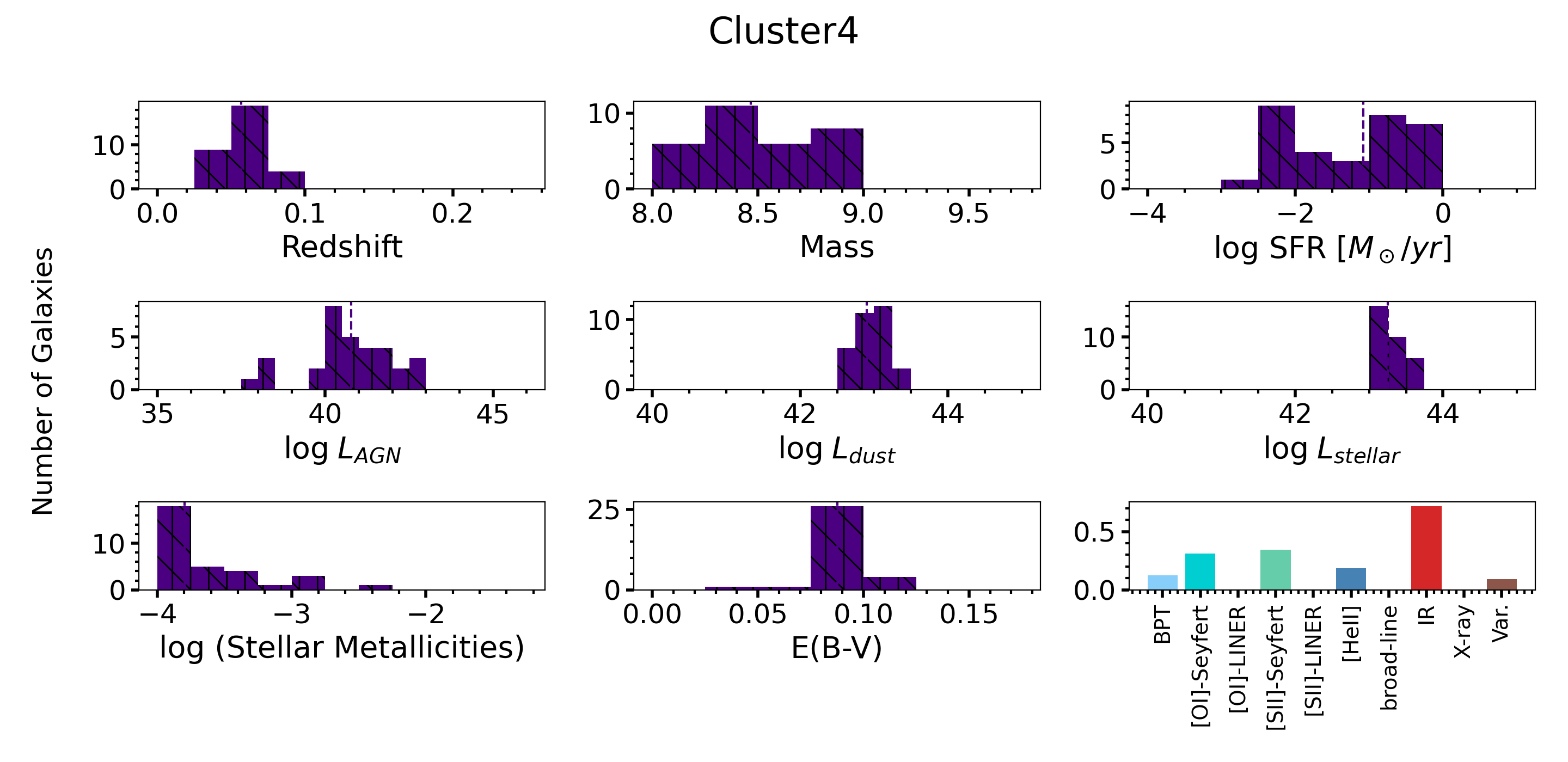} }
    \subfigure{\includegraphics[width=0.742\textwidth]{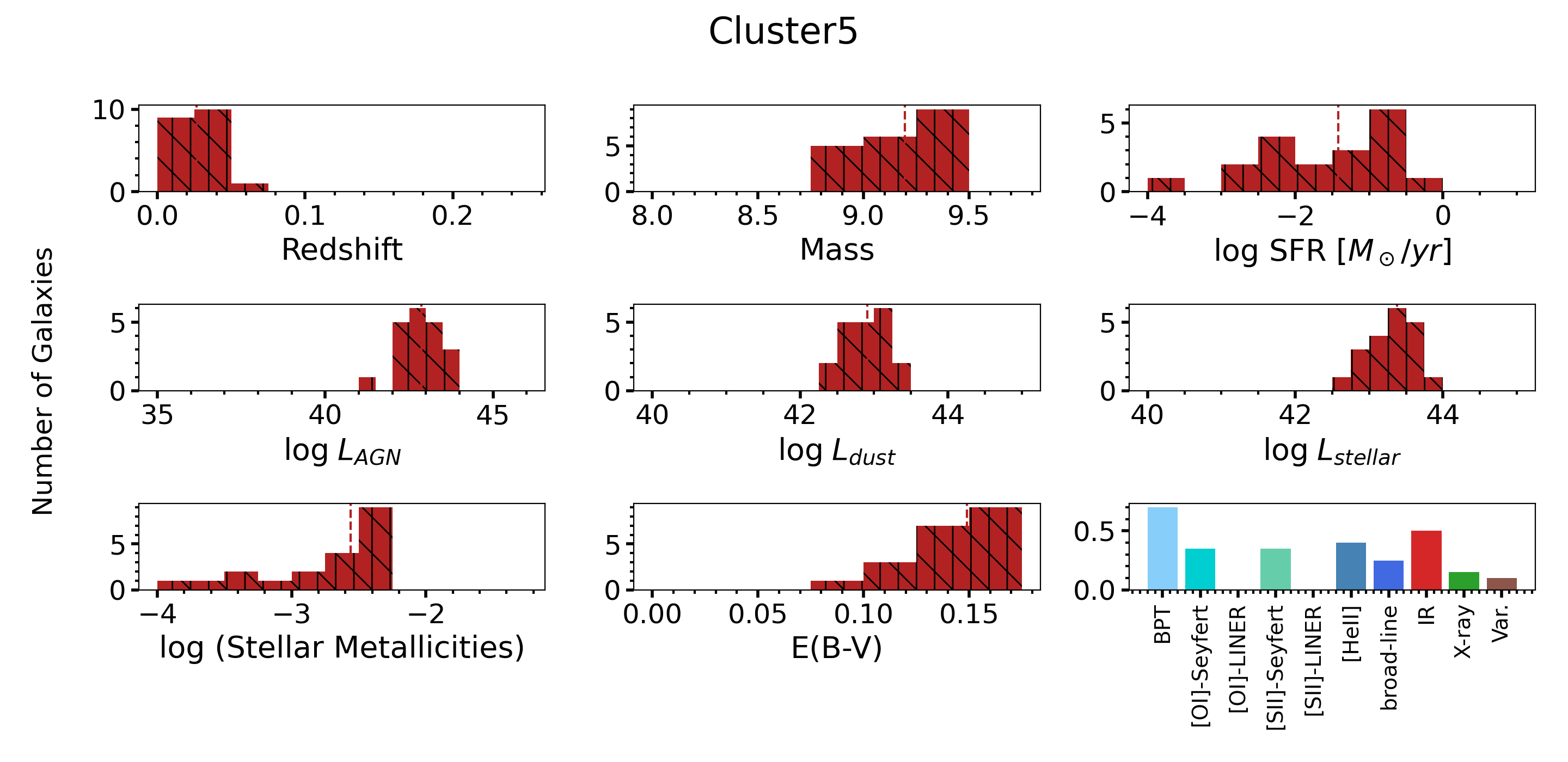} }
    \subfigure{\includegraphics[width=0.742\textwidth]{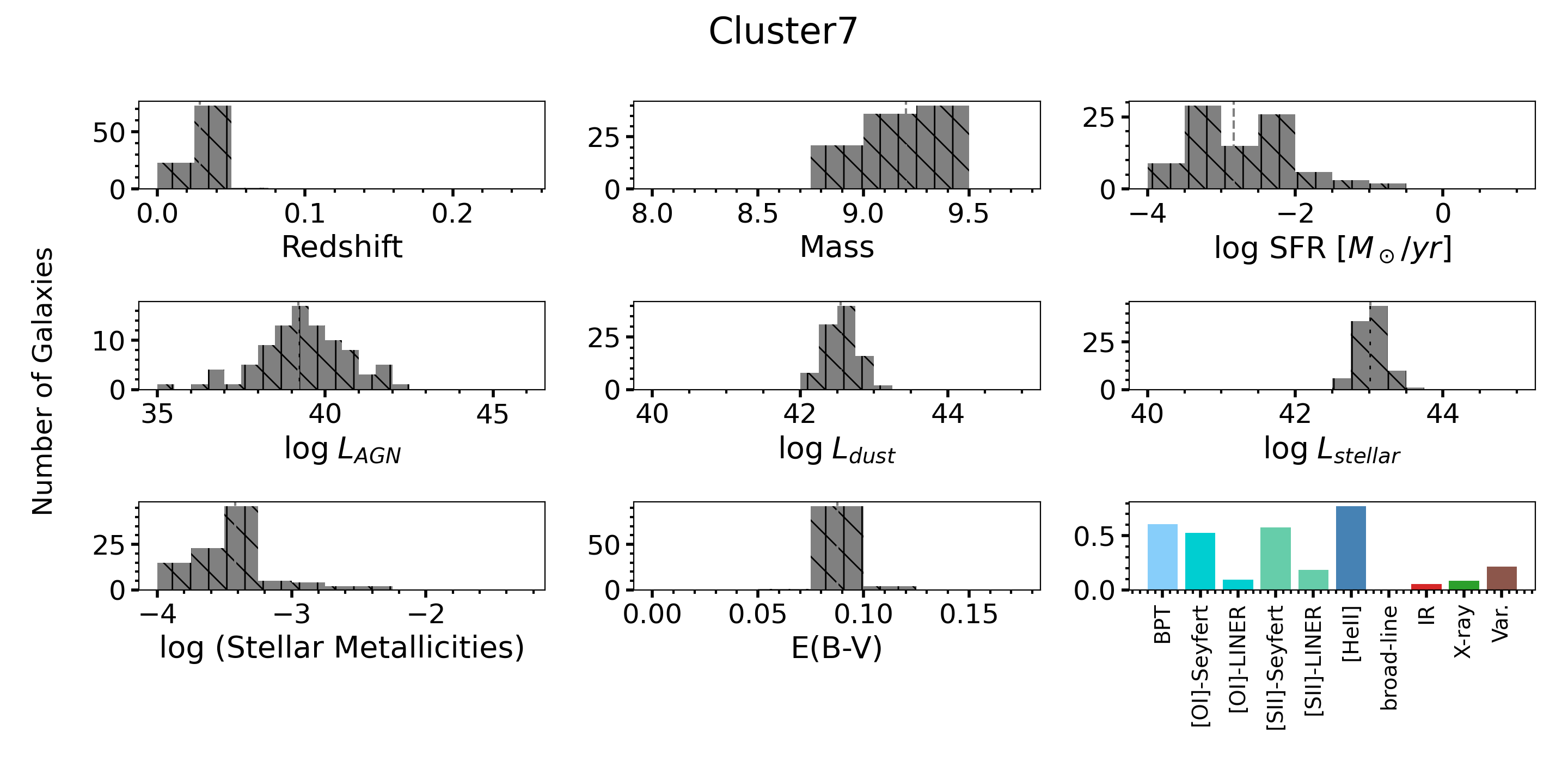} }
            \caption{ Distribution of host galaxy parameters for Clusters 4, 5, and 7 from the t-SNE dimensionality reduction.}
    \label{fig:cluster4-5 7 analysis}
\end{figure*}
\begin{figure*}[h]
    \centering
    \subfigure{\includegraphics[width=0.742\textwidth]{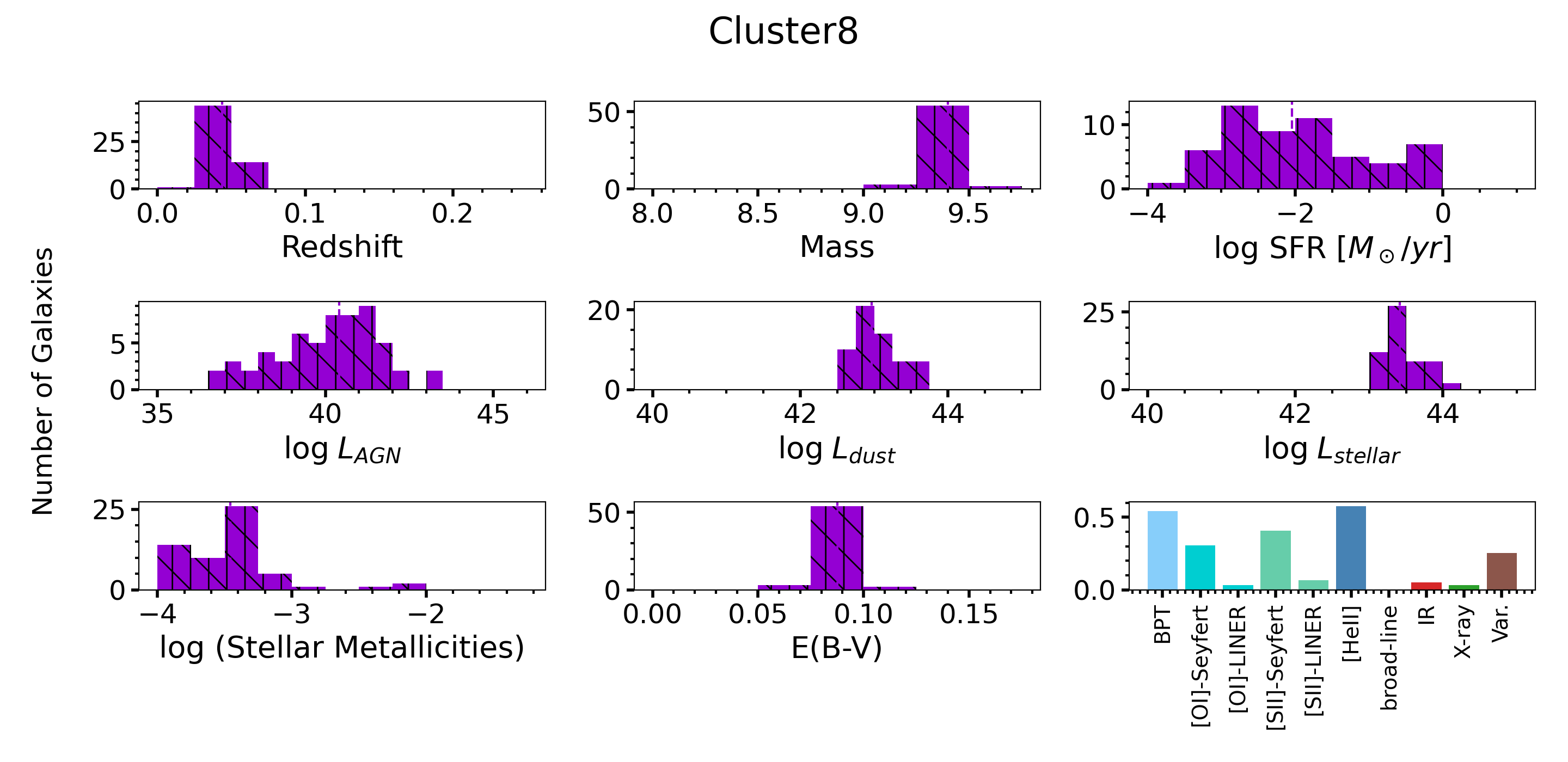} }
    \subfigure{\includegraphics[width=0.742\textwidth]{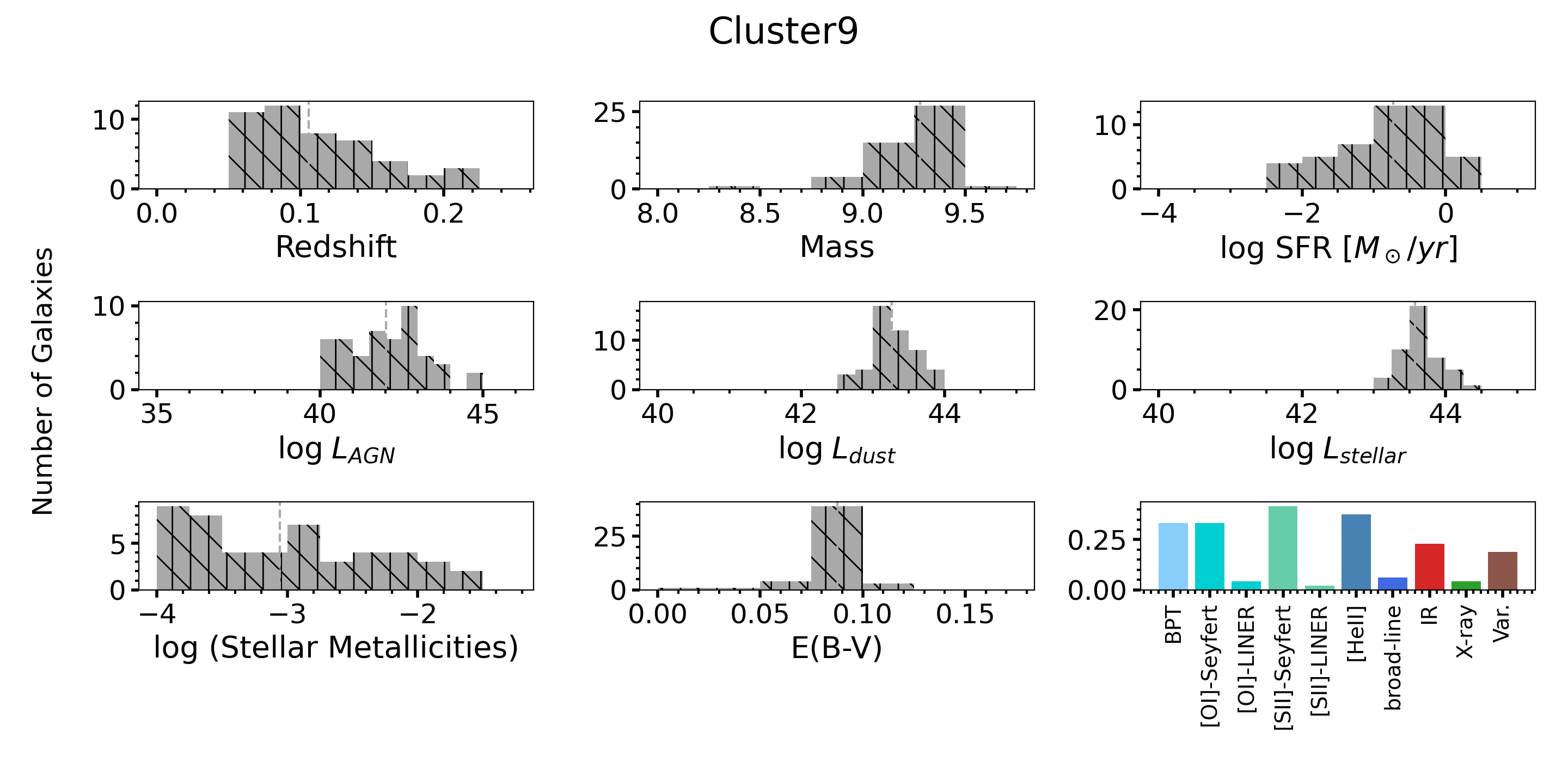} }
            \caption{ Distribution of host galaxy parameters for Clusters 8 and 9 from the t-SNE dimensionality reduction.}
    \label{fig:cluster8-9 analysis}
\end{figure*}
% % % % % % % % % % % %

\section{Percentile Plots of Each Parameter}
Below we give the remaining percentile plots to show the shape of each clusters distributions within the t-SNE input parameters. 

\begin{figure*}[h]%
    \centering
    \subfigure{{\includegraphics[width=0.475\textwidth]{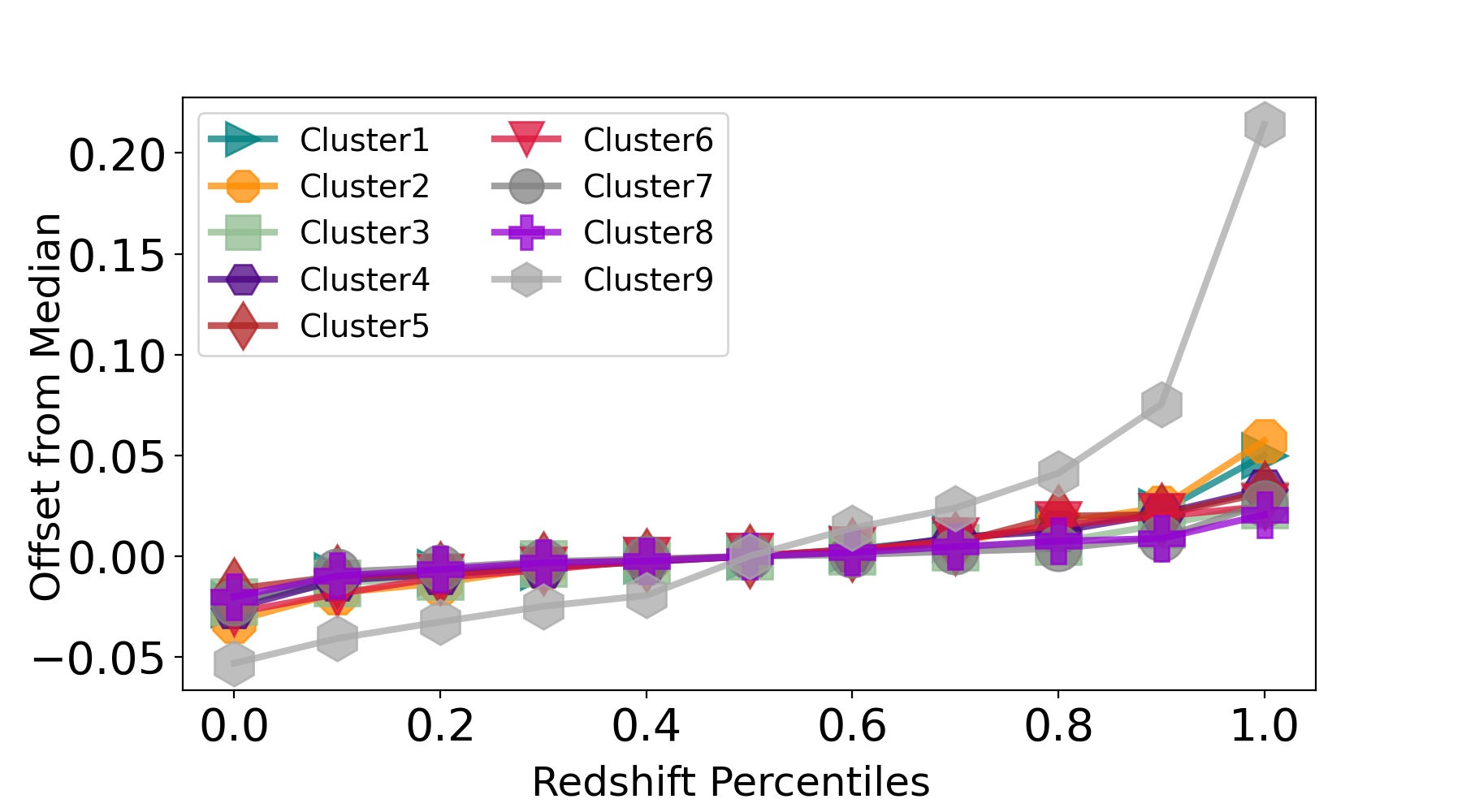} }}%
    \qquad
    \subfigure{{\includegraphics[width=0.475\textwidth]{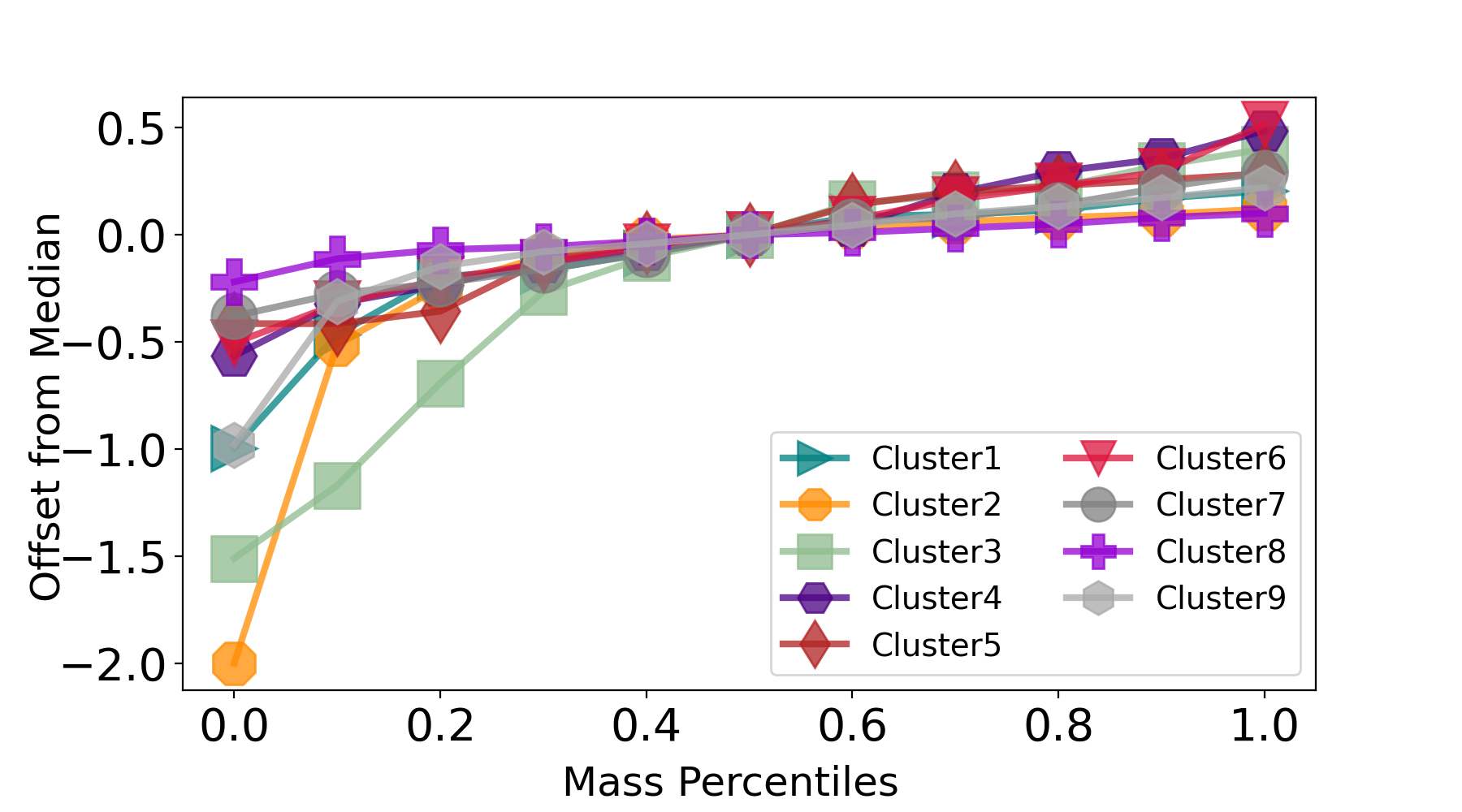} }}%
    %\caption{2 Figures side by side}%
    \caption{Percentile of the distribution vs. offset from median for each cluster within the t-SNE analysis for redshift and mass.}
    \label{fig:Z, Mass percentile}%
\end{figure*}

\begin{figure*}[h]%
    \centering
    \subfigure{{\includegraphics[width=0.475\textwidth]{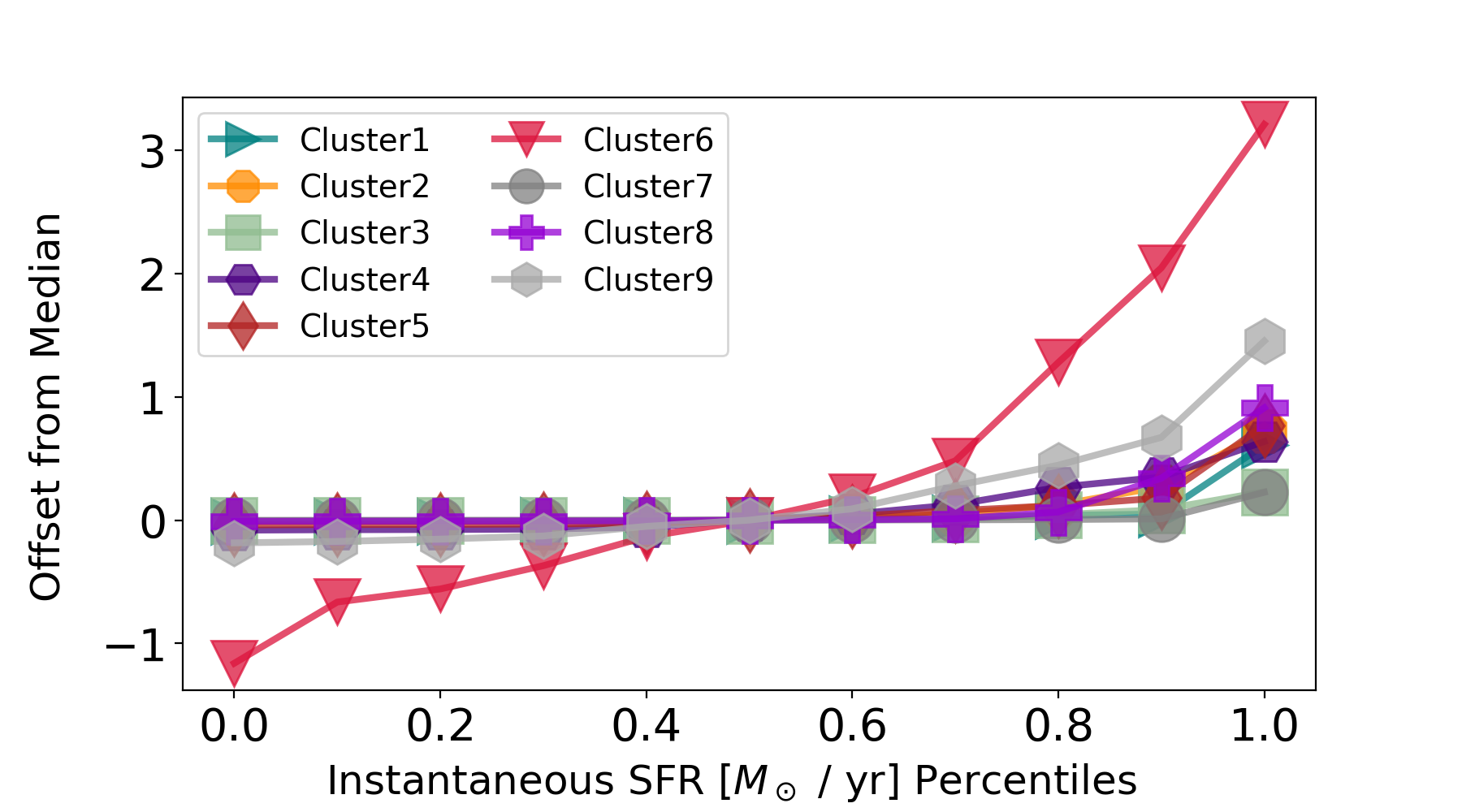} }}%
    \qquad
    \subfigure{{\includegraphics[width=0.475\textwidth]{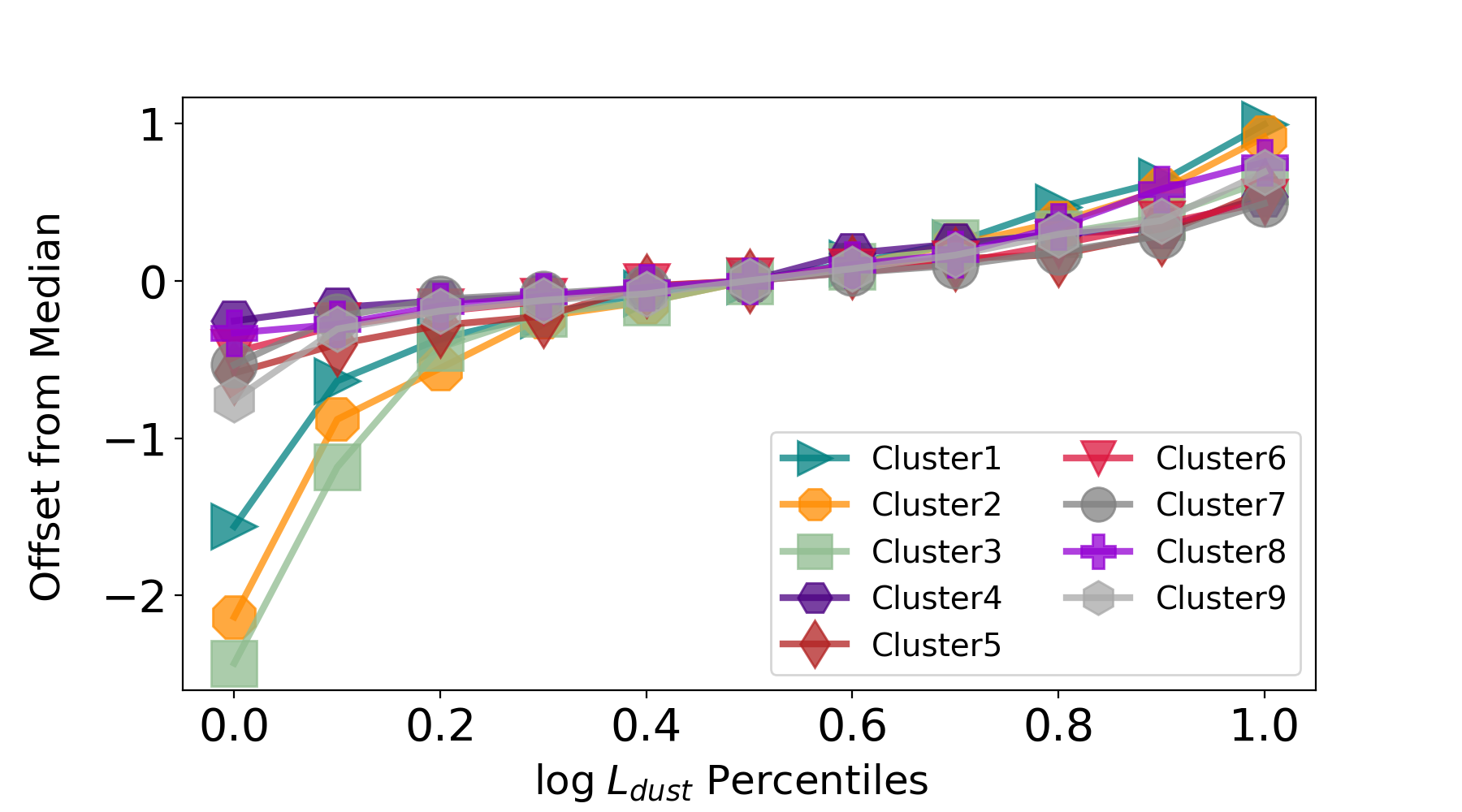} }}%
    %\caption{2 Figures side by side}%
    \caption{Percentile of the distribution vs. offset from median for each cluster within the t-SNE analysis for SFR and $L_{\text{dust}}$.}
    \label{fig: SFR, dust percentile}%
\end{figure*}

\begin{figure*}[h]%
    \centering
    \subfigure{{\includegraphics[width=0.475\textwidth]{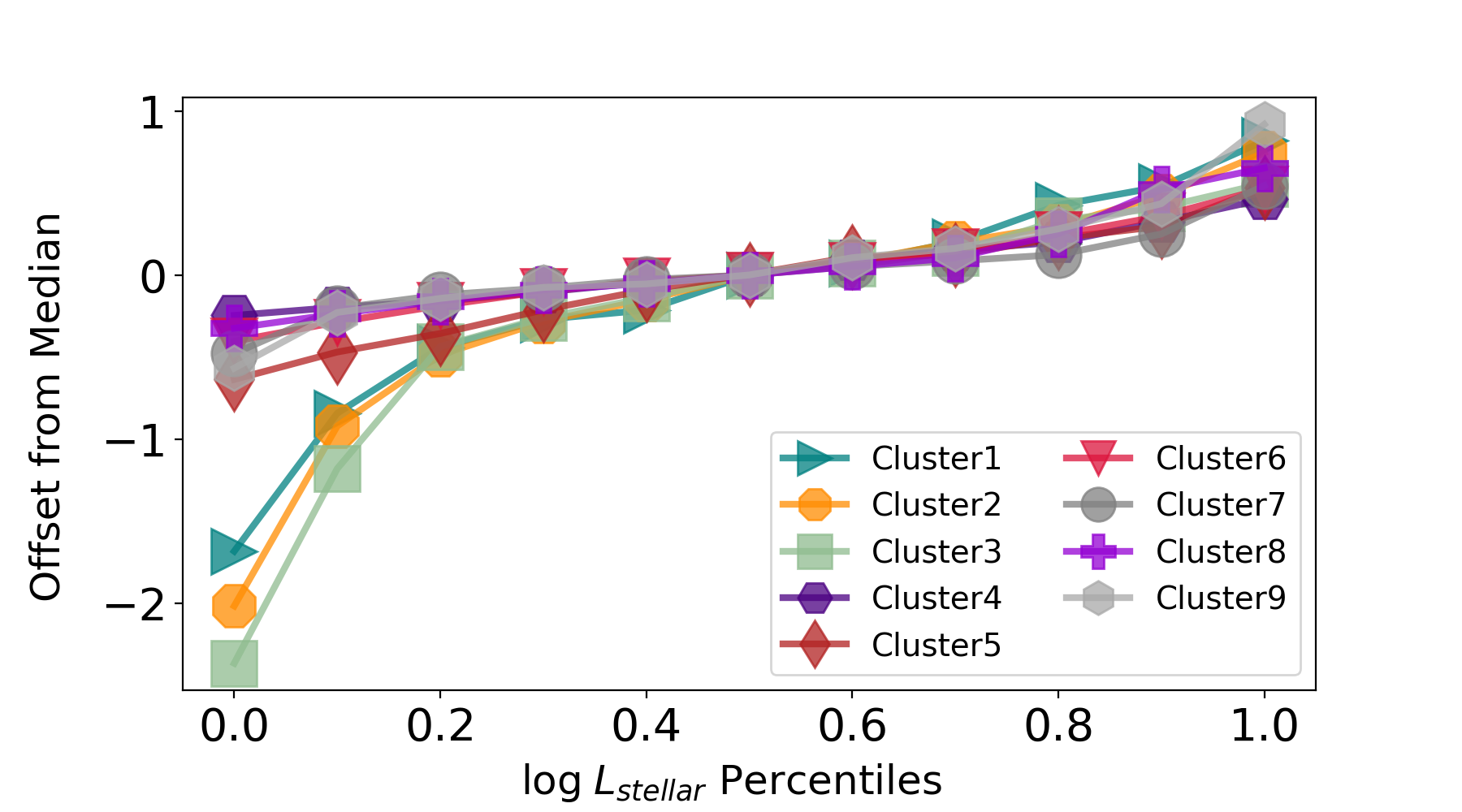} }}%
    \qquad
    \subfigure{{\includegraphics[width=0.475\textwidth]{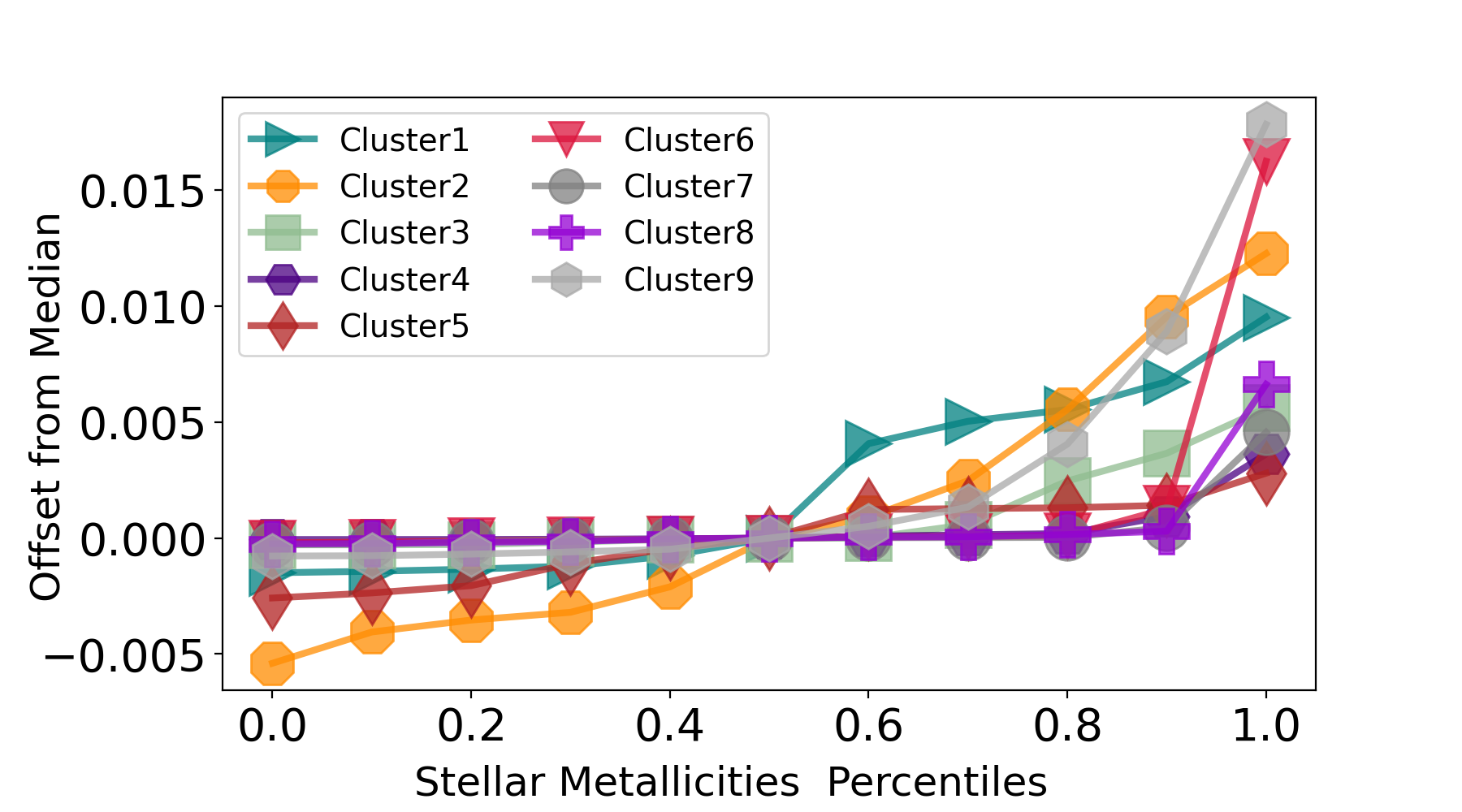} }}%
    %\caption{2 Figures side by side}%
        \caption{Percentile of the distribution vs. offset from median for each cluster within the t-SNE analysis for $L_{\text{stellar}}$ and stellar metallicities.}
    \label{fig:Stellar, Metal percentile}%
\end{figure*}

\begin{figure*}[h]%
    \centering
    \subfigure{{\includegraphics[width=0.475\textwidth]{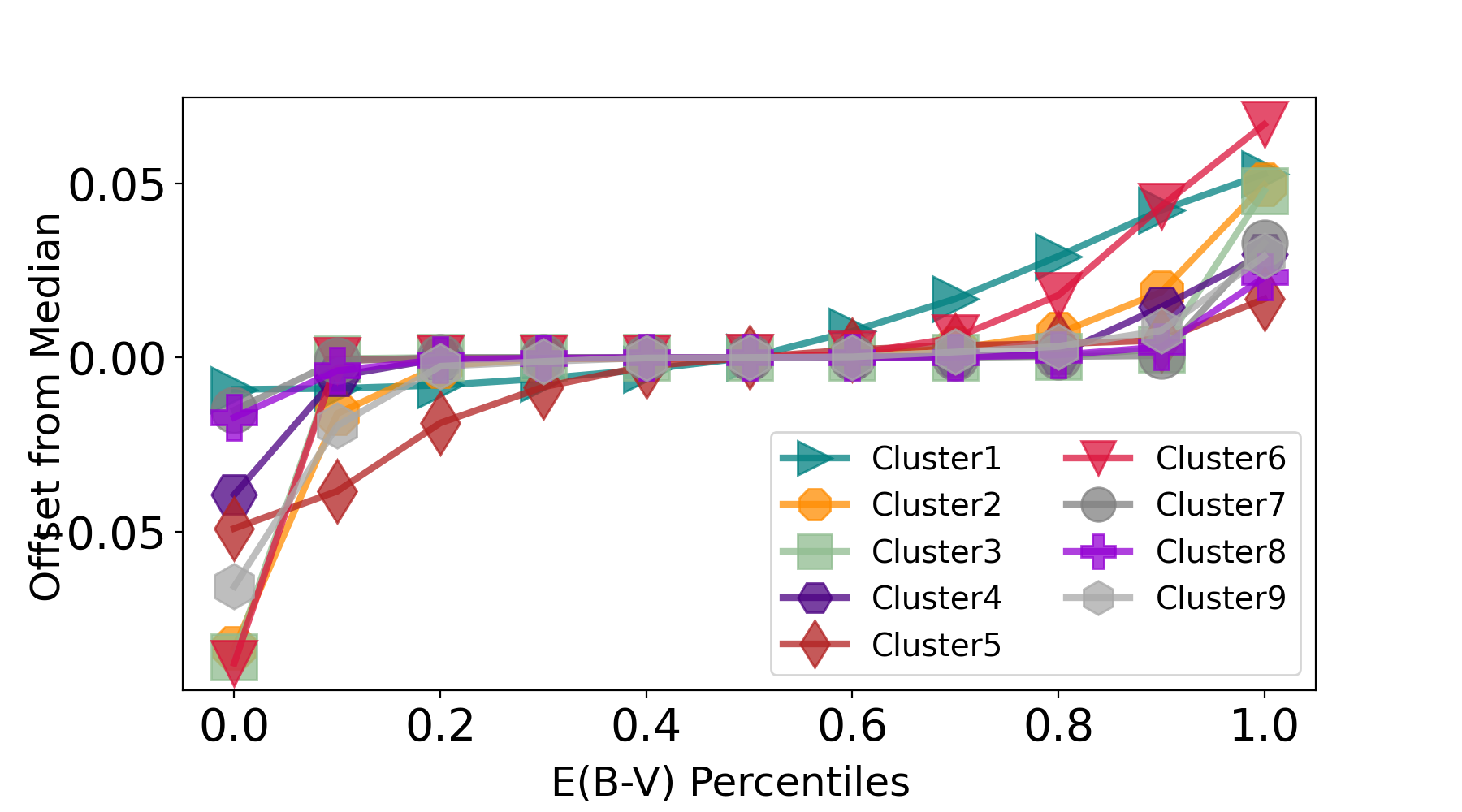} }}%
    \qquad
    %\subfigure{{\includegraphics[width=0.475\textwidth]{Mass_percentiles.png} }}%
    %\caption{2 Figures side by side}%
            \caption{Percentile of the distribution vs. offset from median for each cluster within the t-SNE analysis for excess color.}
    \label{fig: E(B-V) percentile}%
\end{figure*}

\end{document}